\documentclass[letterpaper, aps, prd, twocolumn, superscriptaddress, showpacs, nofootinbib]{revtex4}
\pdfoutput=1

\usepackage{graphicx}
\usepackage{bm}
\usepackage{dcolumn}
\usepackage{color}
\usepackage[sort&compress]{natbib}
\usepackage[latin9]{inputenc}
\usepackage{url}
\usepackage{subfigure}
\usepackage{float}
\usepackage{longtable}
\usepackage{amssymb}
\usepackage{amsmath}
\usepackage{amsfonts}
\newcommand{\Msun}{M_{\odot}}

\begin{document}

\title{Inferring the core-collapse supernova explosion mechanism with
gravitational waves}

\newcommand*{\glasgow}{University of Glasgow, Physics and Astronomy, Kelvin
Building, Glasgow, Lanarkshire G12 8QQ}
\affiliation{\glasgow}

\newcommand*{\caltech}{LIGO 100-36, California Institute of Technology, Pasadena, CA 91125, USA}
\affiliation{\caltech}

\author{Jade Powell}	\affiliation{\glasgow}
\author{Sarah E. Gossan}\affiliation{\caltech}
\author{Joshua Logue}	\affiliation{\glasgow}
\author{Ik Siong Heng}	\affiliation{\glasgow}

\date{\today}


\begin{abstract} 
A detection of a core-collapse supernova (CCSN) gravitational-wave (GW) signal
with an Advanced LIGO and Virgo detector network may allow us to measure
astrophysical parameters of the dying massive star. GWs are emitted from deep
inside the core and, as such, they are direct probes of the CCSN explosion
mechanism. In this study we show how we can determine the CCSN explosion
mechanism from a GW supernova detection using a combination of principal
component analysis and Bayesian model selection. We use simulations of GW
signals from CCSN exploding via neutrino-driven convection and rapidly-rotating
core collapse. Previous studies have shown that the explosion mechanism can be
determined using one LIGO detector and simulated Gaussian noise. As real GW detector
noise is both non-stationary and non-Gaussian we use real detector noise from a 
network of detectors with a sensitivity altered to match the advanced detectors
design sensitivity. For the first time we carry out a careful selection of the
number of principal components to enhance our model selection capabilities. We
show that with an advanced detector network we can determine if the CCSN
explosion mechanism is neutrino-driven convection for sources in our Galaxy and
rapidly-rotating core collapse for sources out to the Large Magellanic Cloud.

\end{abstract}


\maketitle


\section{Introduction}
\label{sec:intro}

More than eighty years after Baade and Zwicky first proposed that core-collapse
supernovae (CCSNe) are massive stars turning into neutron stars at the end of
their life \cite{baade:34}, the CCSN explosion mechanism is
still not fully understood.

Zero age main sequence (ZAMS) stars with $8\,\Msun < M < 100\,\Msun$
form electron-degenerate cores composed primarily of iron-group nuclei in the final
stages of their nuclear burning. Once the iron core exceeds its
effective Chandrasekhar mass (see, e.g.~\cite{baron:90,bethe:90}), it becomes 
gravitationally unstable. Gravitational collapse continues until the inner core is 
dynamically compressed to nuclear densities. At this point, the equation of 
state (EOS) stiffens, the inner core rebounds (typically referred to as 
\emph{core bounce}), and a shock wave is launched outwards from the outer edge 
of the inner core. The shock suffers energy losses due to dissociation of infalling 
iron-group nuclei and neutrino losses from electron capture in the region behind 
the shock. Yielding to ram pressure due to the outer core, the shock stalls and 
becomes an accretion shock. If the shock is not revived within $\sim0.5-3\,\mathrm{s}$,
accretion onto the protoneutron star will lead to further gravitational collapse
and black hole formation~\cite{oconnor:11}. Understanding how the
stalled shock is revived to explode the dying star, the CCSN explosion
mechanism, is one of the most important challenges in CCSN theory today.

There is little information on the CCSN explosion mechanism to be gleaned from 
electromagnetic (EM) observations, as EM emission from CCSNe occurs in optically
thin regions, far from the central engine, and so only secondary information on
the explosion mechanism is available. Observations of CCSN ejecta and pulsar
kicks are indicative of the multidimensional physical processes
driving the explosion \cite{janka:12a,foglizzo:15}. 

Gravitational waves (GWs) and neutrinos, however, are emitted from deep inside
the core and, as such, they are direct probes of the CCSN explosion mechanism.
While GWs from CCSNe have not yet been directly detected, 
the few neutrinos detected from SN1987A confirmed the above picture of core 
collapse~\cite{hirata:87,bionta:87,schaeffer:87}.

The core collapse of massive stars has been considered as a potential source for the 
Advanced LIGO (aLIGO) and Advanced Virgo (AdVirgo) detectors - see ~\cite
{ott:09, kotake:06} for a historical overview. The aLIGO detectors are laser interferometers with 4km arms 
located in Livingston, Louisiana and Hanford, Washington \cite{aLIGO}. AdVirgo is a 3km Italian 
detector expected to join the aLIGO detector network early in 2017~\cite{AdVirgo}.
Recent work by Gossan~\emph{et al.}~\cite{gossan:16} 
shows that GWs from non-rotating and rotating core
collapse may be observable throughout the Milky Way and the Large Magellanic Cloud
(LMC). The rate for these sources is
low at around $\lesssim 2-3\,$ CCSNe per 100yr \cite{1991ARA&A..29..363V, 1993A&A...273..383C, Alexeyev2002,rates:jaz}. No detections were made in the first targeted 
search for CCSNe GWs \cite{2016arXiv160501785A}.

Numerical simulations have allowed a number
of GW emission processes in CCSNe to be identified, including but not limited
to, rotating core collapse and bounce, rotational instabilities,
neutrino-driven convection, prompt convection in the region behind the 
shock, standing accretion shock instability (SASI), and 
asymmetric neutrino emission~\cite{ott:09}. 
The multi-dimensional processes occurring in CCSNe are incredibly complex, and so
even with state of the art simulations, the stochastic nature
of many GW emission processes (e.g. convection, turbulence)
result in a signal with a stochastic phase that cannot be robustly predicted.
In the absence of a robust method to estimate the signal's phase evolution,
matched filtering (the optimal signal extraction method for known signal
morphology in Gaussian noise \cite{owen:99}) cannot be used. To this end,
it is beneficial to associate proposed explosion mechanisms with a set of 
GW emission processes, such that the broad characteristics of GW signals from
each mechanism can be determined. This will allow the detection of GWs from 
CCSNe to be used to infer the CCSN explosion mechanism. 

The first application of numerical GW waveforms for CCSNe to infer the CCSN
explosion mechanism was carried out by Logue~\emph{et al.}
\cite{logue:12} and considered signals from
neutrino-driven convection \cite{janka:07}, rapidly-rotating core
collapse \cite{burrows:07b}, and protoneutron star
pulsations \cite{burrows:06,burrows:07a} (denoted the neutrino mechanism, 
magnetorotational mechanism, and acoustic mechanism, respectively). We hereafter refer to 
this paper as L12. Following previous work by Heng \cite{heng:09} and
R{\"o}ver~\emph{et al.} \cite{roever:09}, principal component analysis
(PCA) \cite{mardia:79} was used to create principal component basis
vectors (PCs) from the GW
signals associated with each explosion mechanism. Linearly polarized
signals were added, or ``injected", in Gaussian noise with the design
sensitivity power spectral density (PSD) of the aLIGO
detectors \cite{aLIGOdesign}, and a Bayesian nested sampling algorithm
\cite{skilling:04} was used to compute the evidence that the signals 
were best represented by each PC basis. Comparison of the evidences for a given
signal then permitted the most likely explosion mechanism to be identified via 
model selection. The algorithm used for this analysis was denoted the Supernova
Model Evidence Extractor (SMEE). 

There were several major limitations to the
SMEE analysis. Firstly, signals were injected into data for one detector,
assuming optimal orientation and sky location for maximal antenna sensitivity of
the detector (see \cite{gossan:16} for information on the antenna
sensitivity of an interferometric GW detector).
Given this, the time-varying antenna sensitivity for a given detector was
not taken into account, and hence the antenna sensitivity considered was
artificially optimistic. Additionally, the single detector network chosen did
not account for the multiple GW detectors scheduled to come online during the
advanced detector era, resulting in limited sensitivity. Further to this, only GW
signals extracted from axisymmetric CCSN simulations were considered, resulting
in linearly polarized signals.
However, EM observations suggest that many, if not most, CCSN explosions exhibit 
asymmetric features \cite{wang:08,chornock:11,smith:12a,sinnott:13,boggs:15}. 
The 3D magnetorotational simulations for rapidly rotating progenitors show a dominant GW polarization is expected for the bounce signal. However, 3D neutrino mechanism simulations show that the stochastic nature of the asymmetric flow structures arising from the SASI and convection will lead to unpolarized GWs from CCSNe \cite{ott:13a,kuroda:16,ott:07,shibata:05,scheidegger:08,andresen:16,kuroda:14,kotake:11,kotake:09, yakunin:14}.
Finally, the use of Gaussian noise meant that the effect of noise
transients present in real GW detector noise could not be studied. Despite these
limitations, the SMEE algorithm demonstrated the ability to distinguish
magnetorotational explosions within the Milky Way ($D \leq 10\,\mathrm{kpc}$), 
while neutrino-driven and acoustic explosions could be distinguished for 
sources closer than $D \leq 2\,\mathrm{kpc}$. The goal of 
this second model selection study is to address the shortcomings of
the original SMEE analysis, and to make more accurate statements on the 
ability to infer the CCSN explosion mechanism from GW observations of CCSNe in 
the advanced detector era.

In this paper, we outline the improvements made to the SMEE analysis to address
several of the limitations described previously. We consider a three detector network
with non-Gaussian, non-stationary detector noise, at multiple GPS times to vary the antenna pattern 
sensitivity. We continue to use linearly polarized GW waveform
catalogs to produce the PCs because, at this time, large waveform
catalogs from 3D CCSN simulations do not exist. Further to this, we do not
consider GW signals from the acoustic mechanism in this study, as this is no
longer considered a viable explosion mechanism for CCSNe \cite{weinberg:08}.

This paper is structured as follows: In Sec.~\ref{sec:mechanism}, we provide an
overview of the CCSN explosion mechanisms considered, the
characteristics of the associated GW emission, and the GW waveform catalogs
chosen for this study. In Sec.~\ref{sec:smee}, we review the method
used in SMEE for Bayesian model selection via nested sampling. In
Sec.~\ref{sec:analysis}, we provide details of the analysis.  
In Sec.~\ref{sec:results}, we show the results of our study at Galactic and extra Galactic distances.
We summarise and discuss the implications in Sec.~\ref{sec:discussion}.


\section{CORE-COLLAPSE SUPERNOVA EXPLOSION MECHANISMS AND THEIR GRAVITATIONAL WAVE SIGNATURES}
\label{sec:mechanism}
In this section, we consider the magnetorotational mechanism and the neutrino
mechanism for driving CCSN explosions. We hereafter provide a broad overview of
the physical processes driving the explosion dynamics, in addition to the
characteristic features imprinted on the associated GW emission. We
direct the reader to L12 for a more 
in-depth description of the explosion mechanisms.

\begin{figure}[!t]
\includegraphics[width=\columnwidth]{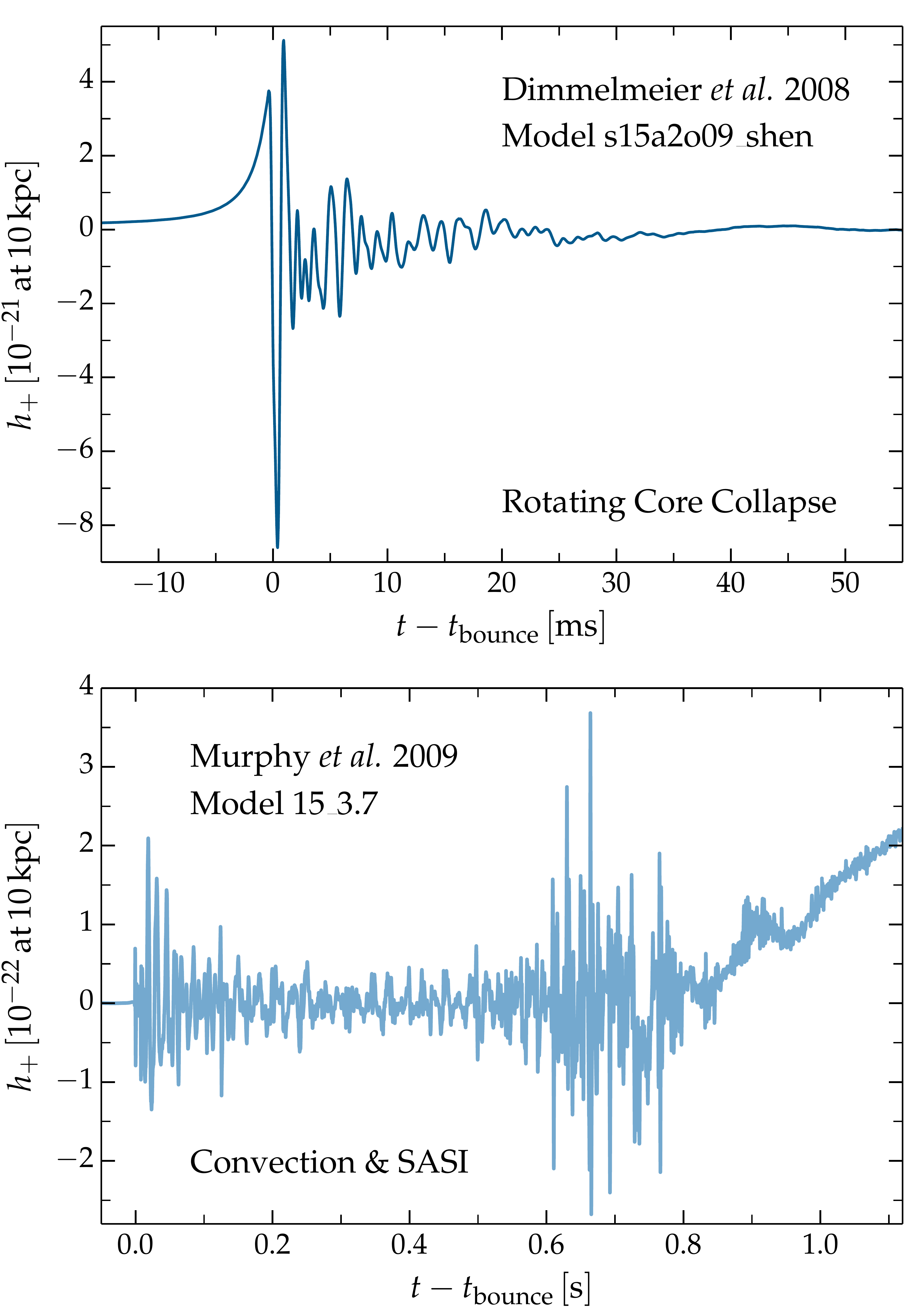}
\caption{Time domain GW strain for representative models of
rotating core collapse (top panel) and neutrino-driven (bottom panel), 
as seen by an equatorial observer at
$10\,\mathrm{kpc}$, drawn from the \texttt{RotCC} and \texttt{C\&S} waveform
catalogs, respectively~\cite{murphy:09,dimmelmeier:08}. We note that 
the typical GW strain from rotating core collapse is 
roughly an order of magnitude larger than the typical GW strain from 
neutrino-driven explosions. In addition, the typical GW signal duration is
roughly an order of magnitude longer for neutrino-driven explosions than for 
rotating core collapse.}
\label{fig:waveforms}
\end{figure}

\subsection{Magnetorotational Mechanism}
\label{subsec:magnetorotational}
Due to conservation of angular momentum, core collapse to a proto-neutron star
results in spin-up of the core by a factor of 
$~1000$~\cite{ott:06spin}. Consequently, rapidly 
rotating pre-collapse cores with periods $\sim1\,\mathrm{s}$ form proto-neutron 
stars with periods on millisecond timescales. Such compact objects have 
rotational energy $\sim10^{52}\,\mathrm{erg}$, a small fraction of which 
could power a strong CCSN explosion, if somehow tapped.

Theory and simulations have shown that magnetorotational processes efficiently extract
rotational energy, and may drive collimated outflows in rapidly rotating core
collapse explosions (see e.g.,~\cite{takiwaki:11,burrows:07b} and references within). Flux
compression due to collapse alone cannot produce the magnetic fields required (of
order $10^{15}\,\mathrm{G}$) for bipolar explosions, given the pre-collapse core
magnetic fields predicted by stellar evolution models~\cite{heger:05}. More plausible
an explanation is magnetic amplification after core bounce, through rotational
winding of the poloidal field into the toroidal field, and the magnetorotational
instability (MRI), though the latter is not well understood in the context of
CCSNe, but some progress towards understanding has been made in recent years \cite{2016MNRAS.460.3316R, 2015Natur.528..376M}. For the 
magnetorotational mechanism to work, simulations suggest that the 
pre-collapse core needs spin period $\lesssim
4-5\,\mathrm{s}$\,\cite{burrows:07b}. 

The GW signal from rapidly rotating CCSNe is dominated by the bounce and
subsequent ring down of the proto-neutron star. Strong centrifugal deformation of
the inner core results from a rapidly rotating pre-collapse core, leading to a
large, time-varying quadrupole moment, which consequently sources a strong burst
of GWs. It is expected that the pre-collapse core angular velocity
distribution is roughly uniform in the inner core \cite{heger:05}, which
is preserved in the subsonically collapsing inner core due to homologous 
collapse. In the supersonically collapsing regions outside the inner core, however, 
homologous collapse drives strong rotation gradients, causing the outer core and region
between the proto-neutron star and the shock to be strongly differentially
rotating \cite{ott:06spin}. Due to this, convection is inhibited in these regions, and
the GW signature of turbulent convection characteristic of non-rotating
core collapse is not present. For slowly rotating core collapse, prompt
convection may contribute to the GW signal on timescales of tens of
ms~\cite{dimmelmeier:08,abdikamalov:14}. Typically, the peak
GW strain from rotating core collapse is $\sim10^{-21} - 10^{-20}$ for 
a source at $10\,\mathrm{kpc}$, and emitted energy in GWs
$(E_{\mathrm{GW}})$ is $\sim10^{-10}-10^{-8}\,M_{\odot}$. The GW energy spectrum is more narrowband 
than for non-rotating core collapse, with most power emitted between
$500-800\,\mathrm{Hz}$, over timescales of a few tens of ms. For pre-collapse
cores with initial spin period less than $\sim0.5-1\,\mathrm{s}$, core bounce
occurs slowly at subnuclear densities, dynamics are dominated by
centrifugal effects, and most energy in GWs is emitted around
$\sim200\,\mathrm{Hz}$ \cite{ott:09,dimmelmeier:08}. It is also possible for 
non-axisymmetric rotational instabilities to develop in the
proto-neutron star, which may source GWs over timescales of hundreds of
ms \cite{ott:07,scheidegger:08,scheidegger:10,shibata:05,rampp:98}.

\subsubsection{GW Waveform Catalogs}
\label{subsubsec:magrotGWcatalog}
For the purposes of this study, we draw from the Dimmelmeier~\emph{et
al.}\,\cite{dimmelmeier:08} waveform catalog to construct the magnetorotational
mechanism PCs, hereafter referred to as the \texttt{RotCC} model. The full
catalog, comprised of 128 waveforms, spans progenitor star ZAMS mass 
($M_{\mathrm{ZAMS}} \in \{12, 15, 20, 40\}\,\Msun$), angular momentum distribution, 
and nuclear matter EOS. The initial angular momentum distribution of the
pre-collapse core is imposed through a parameterized angular velocity profile,
$\Omega_{i}(\bar{\omega})$,
defined as
\begin{align}
\Omega_{i}(\bar{\omega}) &= \frac{\Omega_{c,i}}{1 + (\bar{\omega}/A)^{2}}\,,
\end{align}
where $\bar{\omega}$ is the cylindrical radius, $\Omega_{c,i}$ is the central
angular velocity, and $A$ is the differential rotation length scale. Simulations
are performed across the angular momentum distribution space, considering
strongly differential rotation ($A = 500\,\mathrm{km}$) to almost uniform rotation 
($A = 50000\,\mathrm{km}$); and slowly rotating ($\Omega_{c,i} =
0.45\,\mathrm{rad}\,\mathrm{s}^{-1}$) to rapidly rotating ($\Omega_{c,i} =
13.31\,\mathrm{rad}\,\mathrm{s}^{-1}$) pre-collapse cores. The Lattimer-Swesty 
EOS with incompressibility parameter $K = 180\,\mathrm{MeV}$~\cite{lattimer:91}, 
and Shen EOS with $K = 281\,\mathrm{MeV}$~\cite{shen:98a,shen:98b} are also
used. As the simulations are axisymmetric, the waveform catalog is linearly
polarized. A representative waveform from the Dimmelmeier~\emph{et al.} catalog
is shown in the top panel of Fig.~\ref{fig:waveforms}.
As the main feature of the Dimmelmeier waveforms is the spike at core bounce 
they are still a good approximation of a 3D CCSNe
signal as any rotating model likely stays sufficiently close to
axisymmetry around the bounce and the bounce signal is still clearly
present in 3D magnetorotational waveforms. 

\begin{figure}[!h]
\begin{centering}
\includegraphics[width=\columnwidth]{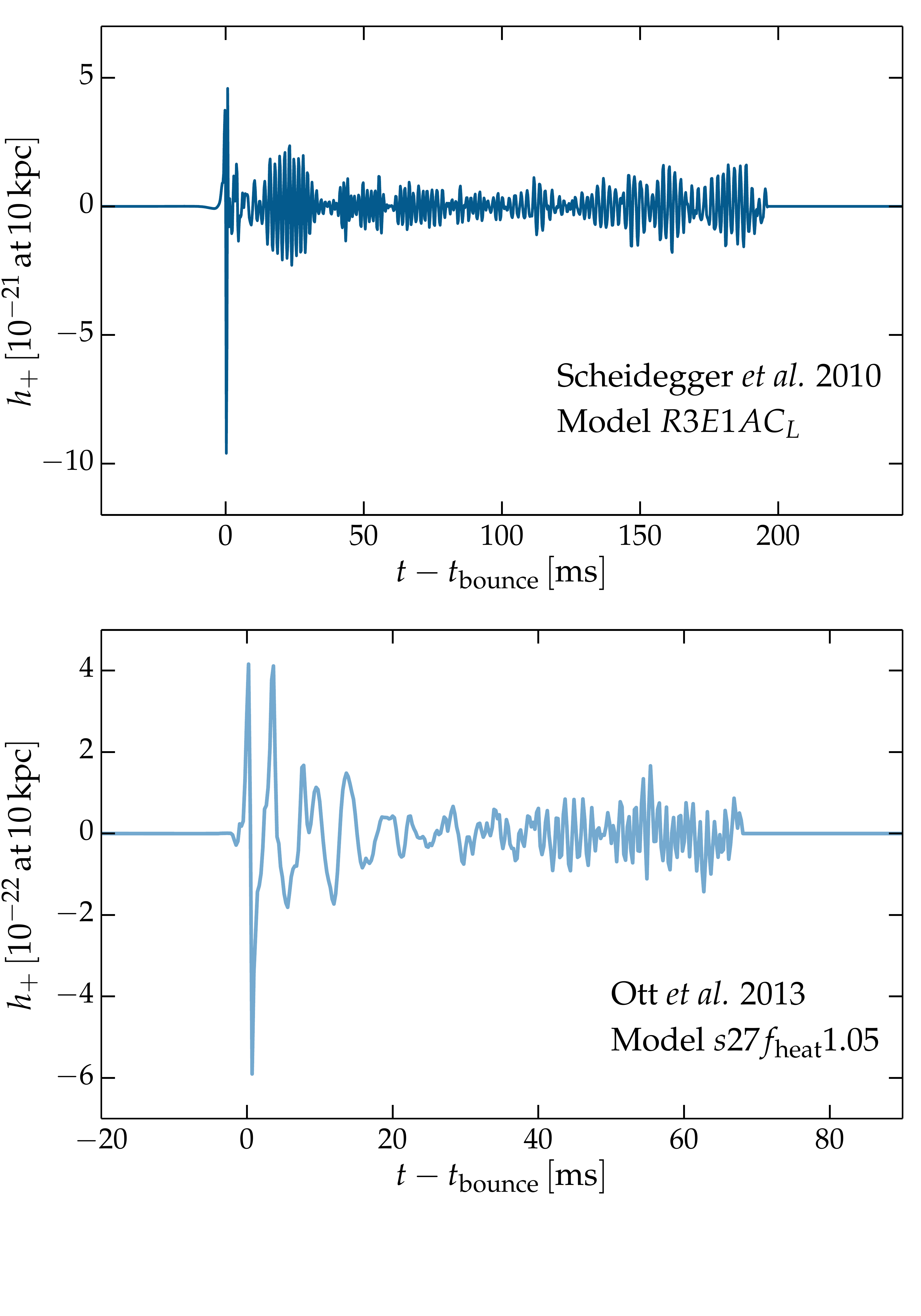}
\caption{Time domain $h_{+}$ GW strain for representative 3D models of
rotating core collapse (top panel) and neutrino-driven convection (bottom panel),
as seen by an equatorial observer at $10\,$kpc, drawn from the Scheidegger~\emph{et al.}~\cite{scheidegger:10b}
and Ott~\emph{et al.}\,\cite{ott:13a} waveform catalogs. }
\label{fig:waveforms_3d}
\end{centering}
\end{figure}

Sample waveforms are also drawn from the following rotating core-collapse simulations:
\begin{itemize}
\item Scheidegger~\emph{et al.}\,\cite{scheidegger:10b} carried out 3D 
magnetohydrodynamical simulations, using a leakage scheme for neutrino
transport. These were performed with a $15\,\Msun$ progenitor
star, and the Lattimer-Swesty EOS with $K = 180\,\mathrm{MeV}$\,\cite{lattimer:91}. 
Due to the 3D nature of the simulations, the Scheidegger~\emph{et al.} 
waveforms have two polarizations. We employ waveforms from models
$\mathrm{R3E1AC}_{\mathrm{L}}$ (moderate pre-collapse rotation, toroidal/poloidal
magnetic field strength of $10^{6}\,\mathrm{G}$/$10^{9}\,\mathrm{G}$)
shown in Fig. \ref{fig:waveforms_3d}, and
$\mathrm{R4E1FC}_{\mathrm{L}}$ (rapid pre-collapse rotation, toroidal/poloidal
magnetic field strength of $10^{12}\,\mathrm{G}$/$10^{9}\,\mathrm{G}$). We
hereafter refer to these waveforms as \texttt{sch1} and \texttt{sch2},
respectively.

\item Abdikamalov~\emph{et al.}\,\cite{abdikamalov:14} performed axisymmetric
general-relativistic hydrodynamics simulations. A $15\,\Msun$ progenitor
star was used, and the Lattimer-Swesty EOS with $K = 220\,\mathrm{MeV}$ 
employed\,\cite{lattimer:91}. We use waveforms from models 
$A1O14$ ($A = 300\,\mathrm{km}$; $\Omega_{c} =
14\,\mathrm{rad}\,\mathrm{s}^{-1}$), $A3O09$ ($A = 634\,\mathrm{km}$;
$\Omega_{c} = 9\,\mathrm{rad}\,\mathrm{s}^{-1}$), and $A4O01$ ($A =
1268\,\mathrm{km}$; $\Omega_{c} = 1\,\mathrm{rad}\,\mathrm{s}^{-1}$. We
hereafter refer to these waveforms as \texttt{abd1}, \texttt{abd2}, and
\texttt{abd3}, respectively.

\end{itemize}

\subsection{Neutrino Mechanism}
\label{subsec:neutrino}
During the collapse of the iron core, and subsequent evolution of the
proto-neutron star to a cold neutron star, approximately
$3\times10^{53}\,\mathrm{erg}$ of energy is released. $99\%$ of this energy is
transported out by neutrinos \cite{bethe:90}. The neutrino mechanism theorizes
that if some small fraction of the energy emitted in neutrinos was
reabsorbed behind the stalled accretion shock, shock heating could re-energize
the shock and drive an explosion. In its early form, the neutrino mechanism was 
first proposed by Arnett \cite{arnett:66}, and Colgate and 
White \cite{colgate:66}, while the modern form of the mechanism was put forward
by Bethe and Wilson \cite{bethe:85}.

The GW signal from neutrino-driven CCSNe is dominated by contributions from
turbulent convection and the standing-accretion-shock instability
(SASI)\,\cite{blondin:03,foglizzo:07,foglizzo:06,scheck:08,kuroda:16}. Immediately after
bounce, there is a burst of GWs from prompt convection \cite{ott:09}, which 
is driven by the negative entropy gradient set up by the stalled shock. The GW
emission dies down as the entropy gradient smooths out, but strengthens as
the SASI becomes nonlinear on timescales of several $100\,\mathrm{ms}$.
Accretion plumes are rapidly decelerated as they enter the region behind the
shock, leading to numerous spikes in the GW signal\,\cite{murphy:09}. The GW
signal is broadband in frequency, with most emission between $100 -
1100\,\mathrm{Hz}$. The signal typically lasts from $\sim0.3-2\,\mathrm{s}$,
with strain $\sim10^{-22}$ for a source at $10\,\mathrm{kpc}$. The total
$E_{\mathrm{GW}}$ from neutrino-driven explosions are of order
$10^{-11} - 10^{-9}\,M_{\odot}$.  

\subsubsection{GW Waveform Catalogs}
\label{subsubsec:neutrinoGWcatalog}
In this study, we use the waveform catalog from 
Murphy~\emph{et al.}~\cite{murphy:09} to construct the neutrino mechanism 
PCs, hereafter referred to as the \texttt{C\&S} model. The 
catalog is comprised of 16 waveforms, extracted using the quadrupole
approximation \cite{mtw}, from axisymmetric Newtonian CCSN simulations.
Electron capture and neutrino leakage are treated using a parameterized scheme,
and only the monopole term of the gravitational potential is included. The 
progenitor models considered are non-rotating, and span the parameter space 
of ZAMS mass ($M_{\mathrm{ZAMS}} = \{12, 15, 20, 40\}\,\Msun$) and total
electron/anti-electron neutrino luminosity. Due to the axisymmetric nature of
the simulations, the waveforms extracted are linearly polarized. While the
nature of the turbulent convection driving such explosions is expected to be 
very different in two- and three-dimensions, broad catalogs of waveforms
extracted from 3D simulations have not yet been produced due to computational
limitations. A representative waveform from the Murphy~\emph{et al.} catalog is
shown in the bottom panel of Fig.~\ref{fig:waveforms}, for a source at
$10\,\mathrm{kpc}$.

Sample waveforms are drawn from the
following non-rotating core collapse simulations:
\begin{itemize}
\item Yakunin~\emph{et al.}\,\cite{yakunin:10} carried out 
axisymmetric Newtonian simulations, using an approximate general relativity 
monopole term of the gravitational potential, and including 
radiation-hydrodynamics. We choose waveforms obtained from the 
simulation of a $15\,M_\odot$ progenitor star referred to as 
\texttt{yak}. Due to
axisymmetry, the extracted waveform is linearly polarized.

\item M\"uller~\emph{et al.}\,\cite{mueller:e12} performed 3D simulations of
neutrino-driven CCSNe with gray neutrino transport and an
inner boundary condition to prescribe the contraction of the
proto-neutron star core. They started the
simulations after core bounce and assumed a time-varying inner
boundary, cutting out much of the proto-neutron star. 
Prompt and proto-neutron star convection only contribute to the
waveforms at late times, and the contraction of the proto-neutron star 
lowers the GW frequency.
As the simulations are 3D, the M\"uller~\emph{et al.}\
waveforms have two polarizations. We use waveform  
models L15-3 and W15-4 (both with a $15\,M_\odot$ progenitor) and model N20-2 (with a $20\,M_\odot$ progenitor) and refer to these waveforms as
\texttt{m\"uller1}, \texttt{m\"uller2} and \texttt{m\"uller3} respectively. 

\item Ott~\emph{et al.}\,\cite{ott:13a} performed 3D
simulations of neutrino-driven CCSNe. The simulations are general-relativistic
and incorporate a three-species neutrino leakage scheme.
As the simulations are 3D, the Ott~\emph{et
al.}\ waveforms have two polarizations, and we use the GW waveform from
model $s27f_{\mathrm{heat}}1.05$ (a $27\,\Msun$ progenitor) shown in Fig. \ref{fig:waveforms_3d}. We hereafter 
refer to this waveform as \texttt{ott}. 
\end{itemize}


\section{SMEE}
\label{sec:smee}
SMEE is designed as a parameter estimation follow up analysis for possible detection
candidates identified by GW burst searches. 
This section gives a brief overview of the Bayesian data
analysis strategy implemented in SMEE. PCA via
Singular Value Decomposition (SVD) is applied to the catalog waveforms to create
signal models that represent each explosion mechanism. Similar techniques have been
used to extract physical parameters of GW signals from binary systems 
\cite{0264-9381-31-19-195010, PhysRevLett.115.121102, 0264-9381-33-8-085003} and in 
characterizing noise sources in GW detectors \cite{powell:15, powell:16}.

In the previous SMEE analysis, a MATLAB implementation
of SMEE was used, which has now been replaced with a faster and more accurate C
implementation that is part of the LIGO data analysis software package LSC
Algorithm Library (LAL) \cite{LAL}.
In particular we use the LALInference package \cite{veitch:15, essick:15, lynch:15}, 
which is designed for parameter estimation of GW signals.

\begin{figure*}[ht]
\begin{centering}
\includegraphics[width=\textwidth, height=13cm]{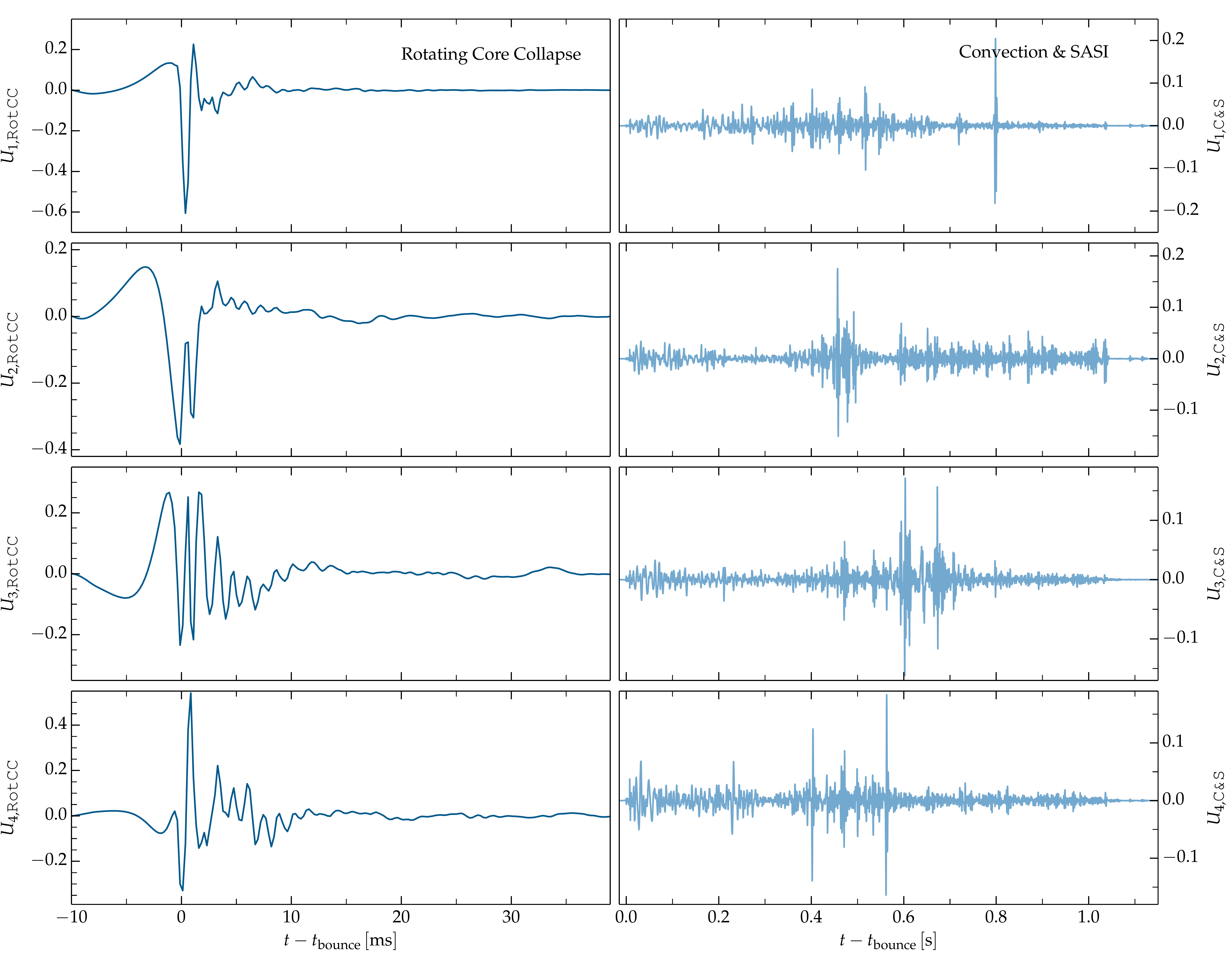}
\caption{(Left) The first four PCs for the \texttt{RotCC} model.
(Right) As for the left, but for the
\texttt{C\&S} model. The first few PCs represent the most common features of the
waveforms used in the analysis. A larger number of PCs is needed to represent the broad set of features in waveforms from the \texttt{C\&S} model. The main feature of the \texttt{RotCC} model PCs is the spike at core bounce. }
\label{fig:1st3PCs}
\end{centering}
\end{figure*}

\subsection{Principal Component Analysis}
\label{subsec:PCA}
PCA can be used to transform waveforms contained in a catalog into a set of
orthogonal basis vectors called principal components (PCs). The first few
PCs represent the main features of a set of waveforms, therefore allowing for a
dimensional reduction of the data set. Before applying PCA waveforms from the 
\texttt{RotCC} model are zero padded and aligned at the spike at core bounce. The 
\texttt{C\&S} waveforms are aligned at the onset of emission.  
By applying SVD to the original data
matrix $D$, where each column corresponds to a supernova waveform, 
the data can be factored such that,
\begin{align}
D &= U\,\Sigma\,V^{\mathrm{T}}\,,
\end{align}
where $U$ and $V$ are matrices whose columns are comprised of the eigenvectors
of $DD^{\mathrm{T}}$ and $D^{\mathrm{T}}D$, respectively. $\Sigma$ is a diagonal
matrix with elements that correspond to the square root of the eigenvalues. The
orthonormal eigenvectors in $U$ are the PCs. As the PCs are ranked by their
corresponding eigenvalues in $\Sigma$, the main features of the data set
are contained in just the first few PCs. Each waveform, $h_{i}$, in the data set can be
reconstructed using a linear combination of the PCs, multiplied by their
corresponding PC coefficients $\beta = \Sigma\,V^{\mathrm{T}}$, such that, 
\begin{align}
h_{i} = A \sum^{k}_{j=1} U_{j}\beta_{j}\,,
\end{align}
where $A$ is an amplitude scale factor and $k$ is the number of PCs. 
Bayesian model selection can then be applied to the signal models. 

\subsection{Bayesian model selection}
Bayesian model selection is used to calculate Bayes
factors that allow us to distinguish between two competing models. The Bayes factor, 
$B_{S,N}$, is given by the ratio of the evidences,
\begin{align}
B_{S,N} = \frac{p(D|M_{S})}{p(D|M_{N})}\,,
\end{align}
where $M_{S}$ and $M_{N}$ are the signal and noise models, respectively. The 
evidence is given by the integral of the likelihood multiplied by the prior 
across all parameter values. 
For a large number of parameters the evidence integral can become difficult.
This problem is solved using a technique known as nested sampling. A detailed
description of nested sampling is given in L12
and elsewhere~\cite{sivia:96,veitch:15}. 

For convenience we take the logarithm of the Bayes factor,
\begin{align}
\log B_{S,N} = \log[p(D|M_{S})] - \log[p(D|M_{N})]\,.
\end{align}
If $\log B_{S,N} > 0$, the signal model is preferred over the noise
model. Conversely, if $\log B_{S,N} < 0$, the noise model is preferred
over the signal model. 
In this same way, the evidence for two different explosion models,
\texttt{RotCC} and \texttt{C\&S}, can be compared as
\begin{align}
\log B_{\texttt{RotCC}-\texttt{C\&S}} = \log B_{\texttt{RotCC},N} - \log B_{\texttt{C\&S},N}.
\end{align}
 
Uniform priors are applied to each PC coefficient, with prior ranges set by the
catalog waveforms padded by $\pm10\%$ to account for uncertainty due to 
the lack of available waveforms, and a uniform-in-volume prior is applied to the
amplitude parameter as it scales with distance. A full description of the likelihoods used is given in L12.

A galactic SN will have coincident EM and neutrino signals, ensuring that the
sky location of the target source will be known. Online searches for
GW bursts can also produce sky-maps of the location of the GW
signal \cite{essick:15}. For this reason, we fix the sky
location of the source as a known parameter.

\subsection{Number of PCs}
\label{subsec:pcnum}

Previously in L12, the relative complexity of the \texttt{RotCC} and
\texttt{C\&S} models was not taken into consideration when selecting the 
number of PCs. 

\begin{figure}[!h]
\begin{centering}
\includegraphics[width=\columnwidth]{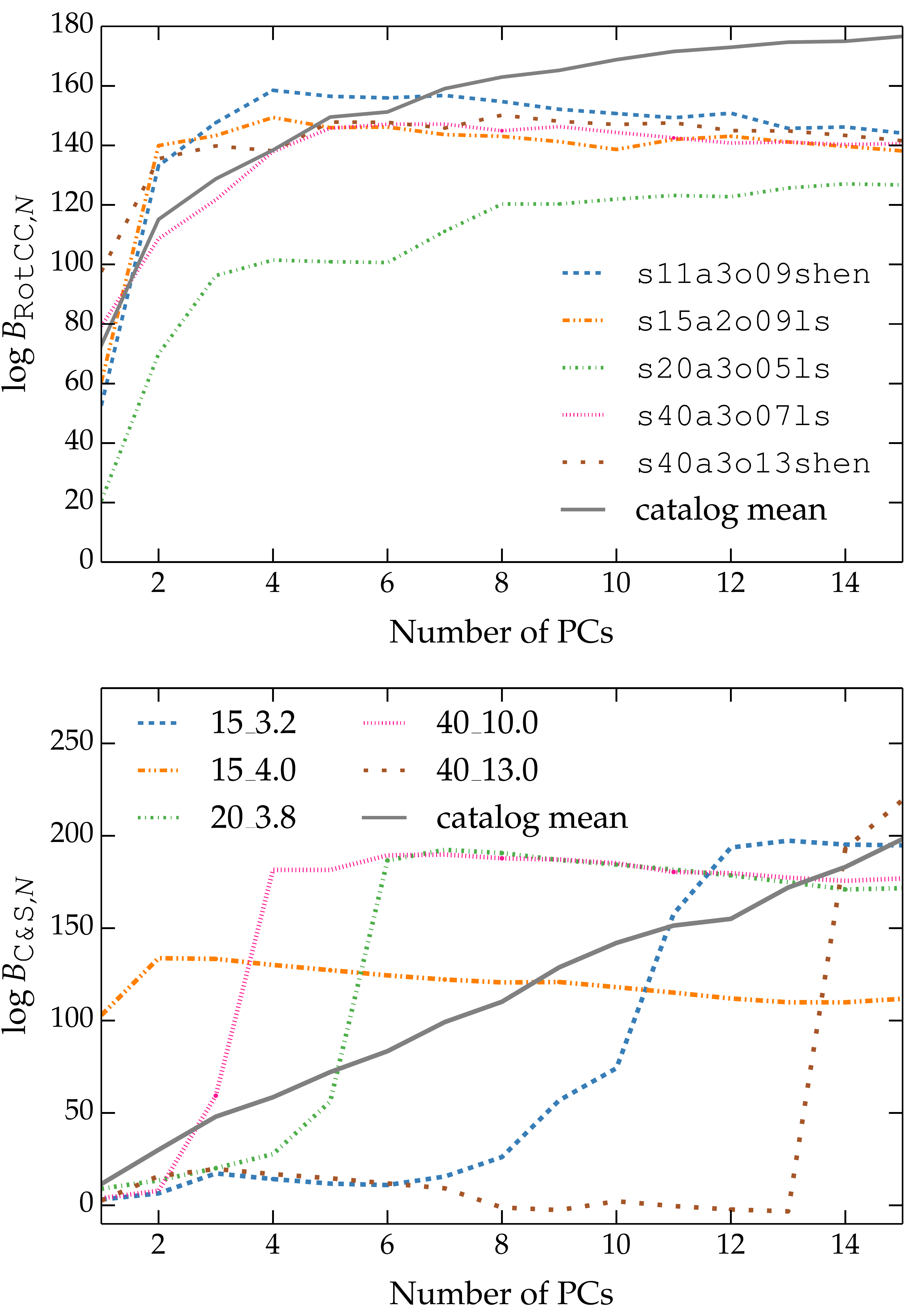}
\caption{The log ${B_{S,N}}$ values for an increasing number of PCs for the
\texttt{RotCC} (top panel) and \texttt{C\&S} (bottom panel) models using five
representative waveforms from each mechanism. Log ${B_{S,N}}$ increases
as more PCs are added and more information is gained about the injected signal.
Log ${B_{S,N}}$ will decrease if the model becomes too complex. }
\label{fig:numPCs}
\end{centering}
\end{figure}

Fig.~\ref{fig:1st3PCs}
shows the first four PCs for the \texttt{RotCC} and \texttt{C\&S} models. 
It is clear that the time domain structure of the \texttt{C\&S} model is far
more complex than that for the \texttt{RotCC} model. Due to this, fewer
\texttt{RotCC} PCs are typically needed to faithfully reconstruct GW signals
from rotating core collapse, than the number of \texttt{C\&S} PCs needed to
reconstruct GW signals from neutrino-driven CCSNe. To account for this, we aim
to quantify the impact of the number of PCs for each model.

This is typically achieved by studying the variance encompassed by each PC, and
using the number of PCs that cumulatively contain above some fraction of the
total variance~\cite{PCA-clustering, powell:15}. However, as this
method only uses the waveforms it does not account for the limitations of
the analysis method implemented in SMEE. Bayesian model selection favours simpler
models and this could increase errors when results are more uncertain, such as when the SNR of the GW 
signal is low \cite{sivia:96}. To this end, we
determine the optimal number of PCs from the behaviour of $\log B_{{S,N}}$ for
both models across the waveform catalogs.

In Fig.~\ref{fig:numPCs} we show $\log{B_{S,N}}$ for five representative waveforms
from the \texttt{RotCC} and \texttt{C\&S} models, which span the
parameter space of the catalogs. We inject all of the signals with an SNR of 20, as $\log{B_{S,N}}$ is
also proportional to the SNR of the signal. Larger SNRs produce larger values of $\log{B_{S,N}}$. 
As the number of PCs is increased the model becomes a better match for the 
signal in the data and $\log{B_{S,N}}$ will increase sharply. After an
ideal number of PCs is reached no further information about the signal is gained by
adding more PCs and $\log{B_{S,N}}$ stops increasing. 
If more PCs are added after the ideal number then $\log{B_{S,N}}$ will begin to decrease 
due to an Occam factor that occurs as the signal model becomes too complex.

The waveforms in the \texttt{RotCC} catalog have a small variance and therefore a small 
number of PCs are needed to represent the entire catalog. The \texttt{C\&S} model 
has greater variance in the catalog waveforms and a larger number of PCs are required 
to accurately represent all the features included in all waveforms. 
We select $6\,$PCs for the \texttt{RotCC} model and $9\,$PCs for the \texttt{C\&S} model 
to maximise the number of features represented in the PCs whilst minimising the penalty that
occurs when the model is too complex or one model is significantly more complex than the other. 

\section{Analysis}
\label{sec:analysis}

In L12 simulated Gaussian noise was considered in a
single aLIGO detector in the context of a sky position where
antenna sensitivity to linearly polarized GW signals was maximised. For this study, 
SMEE has been extended to incorporate a three detector network, which consists of the 
two aLIGO detectors and the AdVirgo detector, hereafter referred to as H1, L1, and V1, respectively.

Real data from GW detectors is non-stationary and non-Gaussian
and, as such, it is important to test 
our analysis in real non-stationary, non-Gaussian noise. We use the
observational data taken by H1 and L1  
during the S5 science run, and data taken by V1 during the VSR1 
science run, which is now publicly available via the LIGO Open Science 
Center (LOSC)~\cite{vallisneri:15}. This data is recolored to the design
sensitivity Power Spectral Density (PSD) of aLIGO and AdVirgo, as outlined
in\,\cite{gossan:16}, which permits a more realistic estimation
of the sensitivity of our analysis in future advanced detector observation runs.
The detectors are expected to reach design sensitivity in 2019.

The antenna response of the detectors is periodic with an associated
timescale of one sidereal day, due to the rotation of the Earth. As a
consequence of this, the sensitivity of any GW analysis using stretches of
data much shorter than this timescale is strongly dependent on
the antenna response of the detectors to the source location at the relevant
GPS time. To represent time-averaged sensitivity of the detector network, we choose 
10 GPS times spread throughout a 24 hour period. 

\section{Results}
\label{sec:results}

\subsection{Response to Noise}

The response of SMEE to instances of simulated Gaussian noise was investigated in L12 
to better understand the results in the presence of real signals. 
As SMEE is now implemented in C, and the relative complexity of the waveforms is now accounted for in 
the number of PCs, we recalculate the noise response using 1000 instances of simulated aLIGO and AdVirgo 
design sensitivity noise. 

\begin{figure}[!h]
\begin{centering}
\includegraphics[width=\columnwidth]{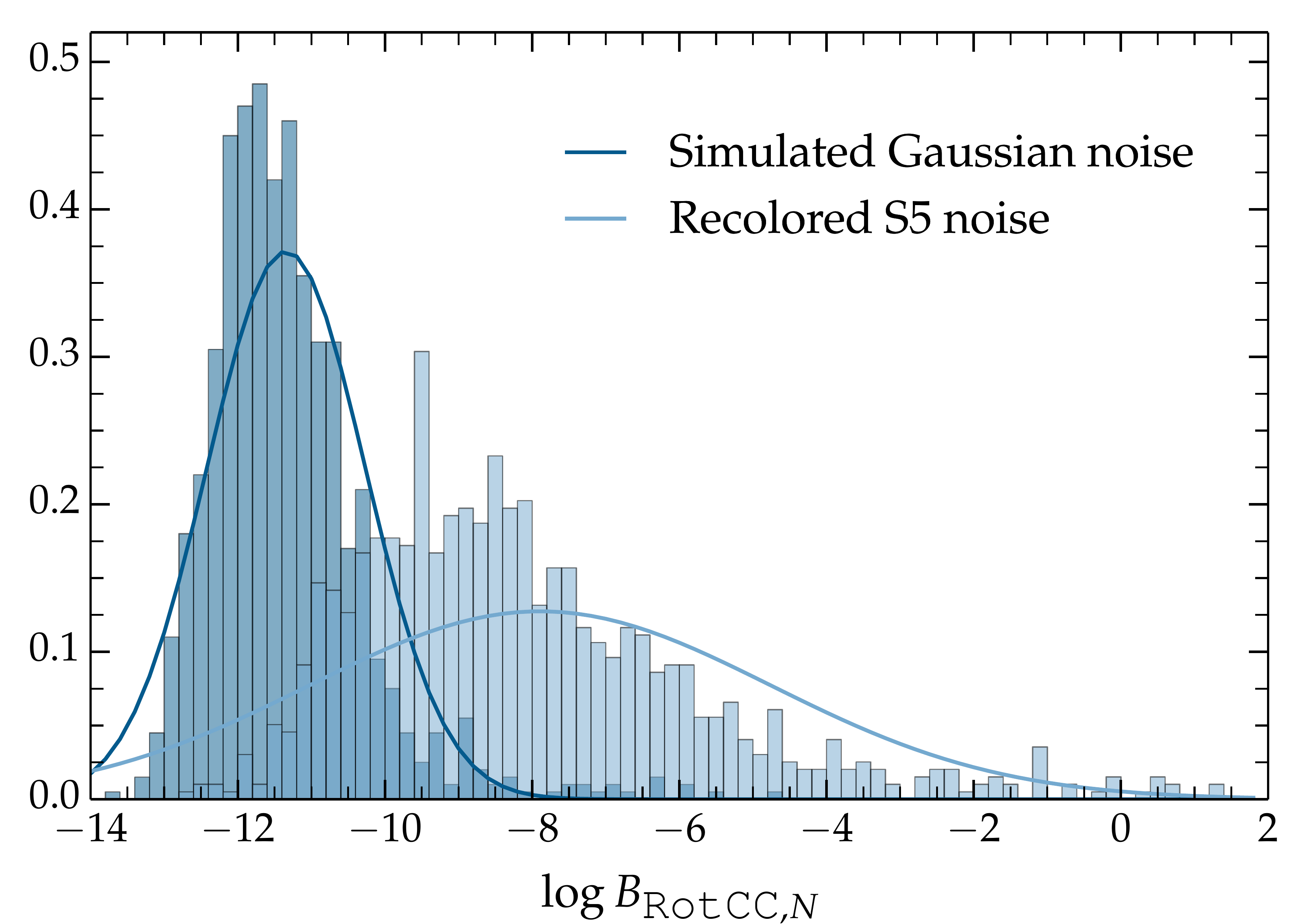}
\includegraphics[width=\columnwidth]{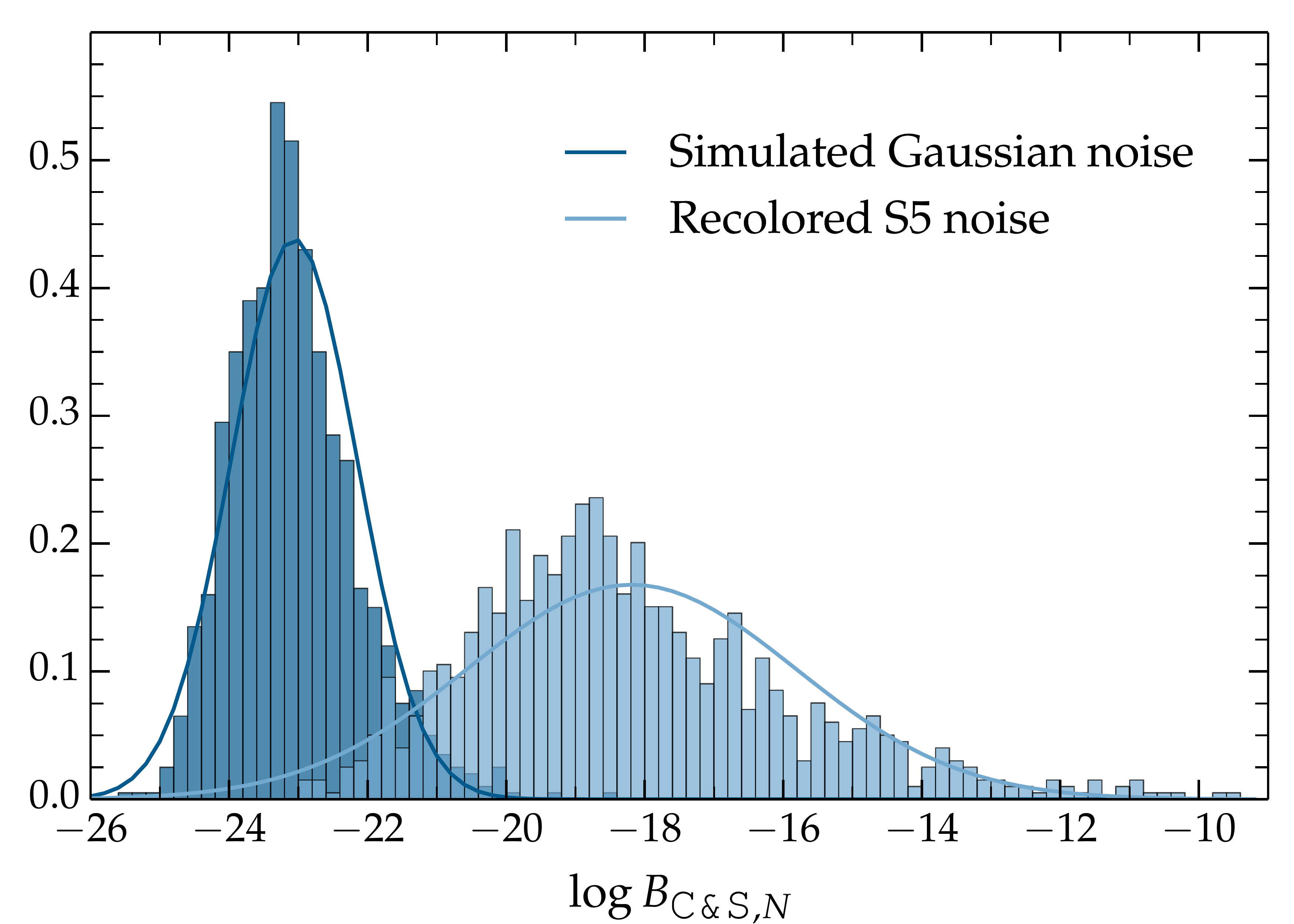}
\caption{Response of SMEE to 1000 instances of simulated and recoloured aLIGO
and AdVirgo design sensitivity noise for \texttt{RotCC} with 6 PCs (top panel) and
\texttt{C\&S} with 9 PCs (bottom panel). Transient noise artefacts and lines in
the real data can increase log $B_{S,N}$ and the standard deviation of the noise
response. }
\label{fig:noiseresp}
\end{centering}
\end{figure}

In Fig.~\ref{fig:noiseresp} $\log B_{S,N}$ for 1000 instances of Gaussian and real
non-stationary, non-Gaussian noise is shown. The $\log B_{S,N}$ values are obtained
by running SMEE on 1000 GPS times, during the 24 hour period of data, which contains no GW signals. 
The mean values are $-12$ for 
the \texttt{RotCC} model and $-23$ for the \texttt{C\&S} model in Gaussian noise and 
$-9$ for the \texttt{RotCC} model and $-19$ for the \texttt{C\&S} model in the recolored noise.
Short duration transient noise artefacts and lines in the data increase SMEE's response to noise and increase
the standard deviation of the noise response.
In L12 a threshold value of $5$ on $\log B_{S,N}$ was set
using the standard deviation of the noise response. We increase the threshold 
on the value of $\log B_{S,N}$ to $10$ to account for the increased variation 
in the noise response found in the real non-Gaussian data.

\subsection{Determining the Core-Collapse Supernova Explosion Mechanism}
\label{subsec:explode}


\begin{figure*}[ht]
\begin{centering}
\subfigure[]{\label{fig:bfhists_a}\includegraphics[width=8.5cm]{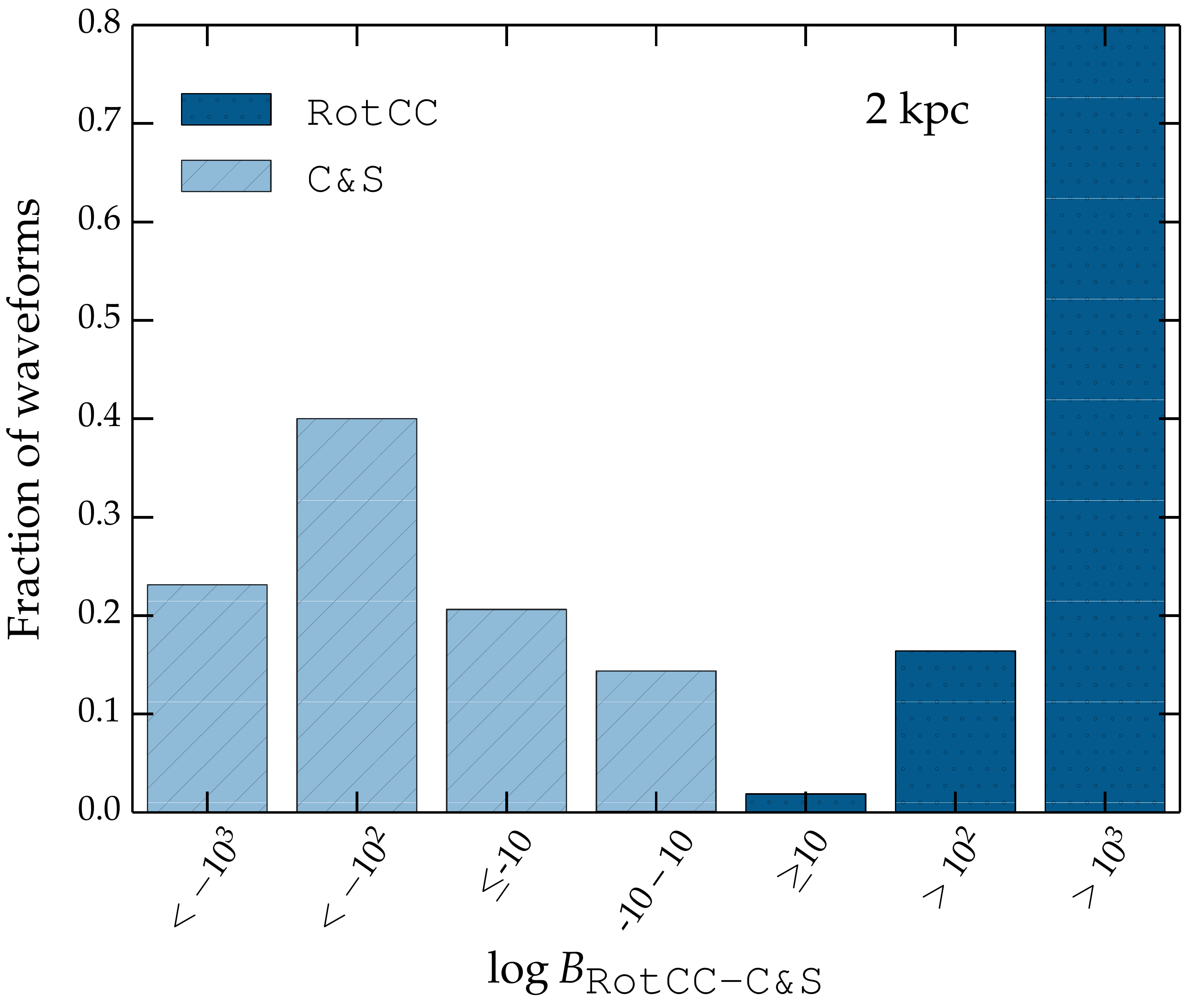}}
\subfigure[]{\label{fig:bfhists_b}\includegraphics[width=8.5cm]{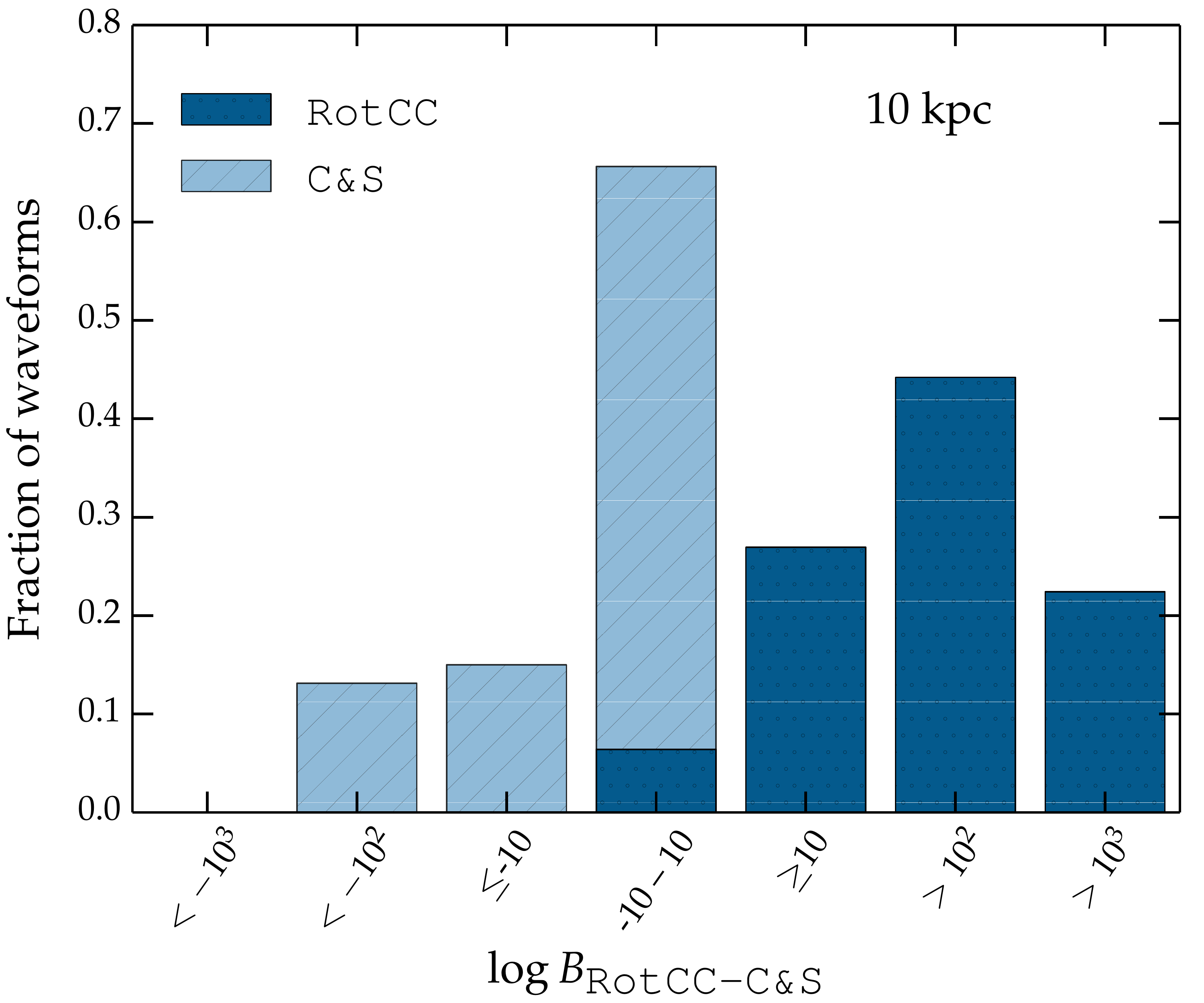}}
\subfigure[]{\label{fig:bfhists_c}\includegraphics[width=8.5cm]{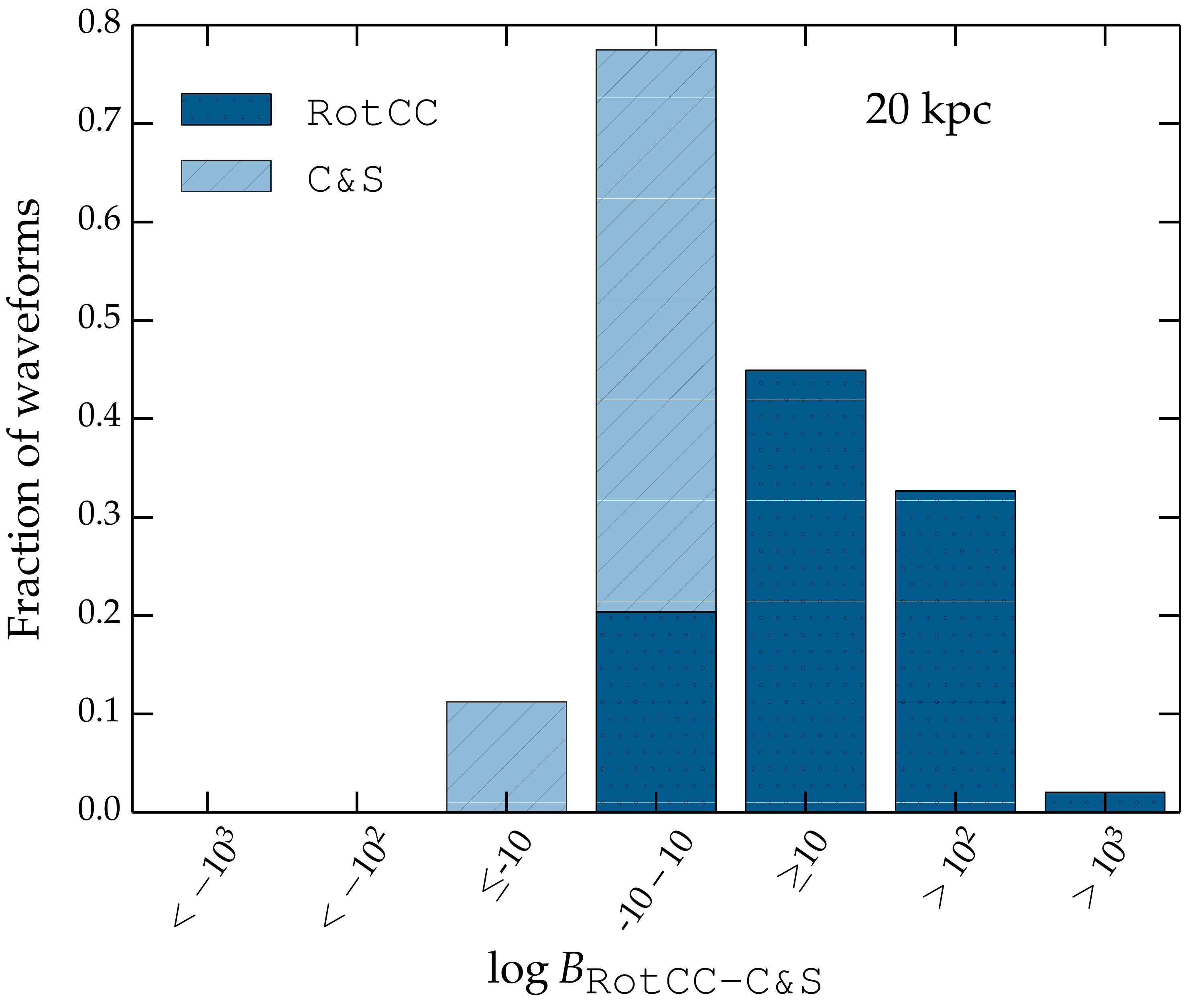}}
\subfigure[]{\label{fig:bfhists_d}\includegraphics[width=8.5cm]{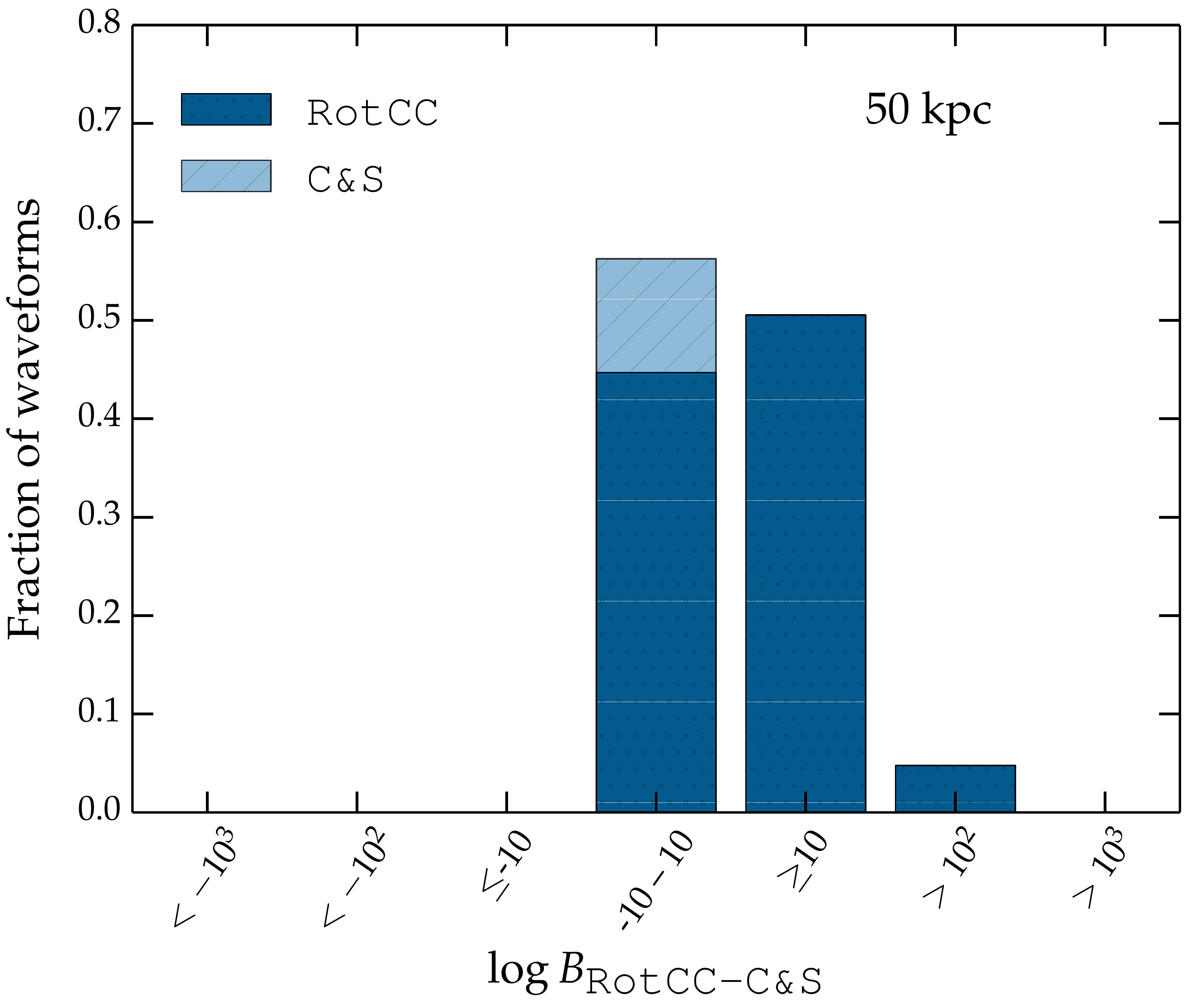}}
\caption{Log $B_{\texttt{RotCC}-\texttt{C\&S}}$ for waveforms injected from the \texttt{RottCC} and \texttt{C\&S} catalogs. (a) At $2\,$kpc and the the sky position of the Galactic center the explosion mechanism is correctly determined for all 1437/1440 detected waveforms. (b) At $10\,$kpc 1198/1440 waveforms are detected and their explosion mechanism is correctly determined. (c) Almost all the \texttt{C\&S} waveforms have an SNR too
small for them to be detected at $20\,$kpc. (d) Distance of $50\,$kpc and sky
position of the Large Magellanic Cloud.}
\label{fig:bfhists}
\end{centering}
\end{figure*}

To test SMEEs ability to determine the explosion mechanism 
all 128 \texttt{RotCC} and 16 \texttt{C\&S} waveforms 
are injected at 10 GPS times giving a total of 1440 injected signals at each distance. The sky position of the
Galactic center is used at distances of $2\,$kpc, $10\,$kpc and
$20\,$kpc to show how well the explosion mechanism can be determined for sources
throughout the Galaxy.

Table.~\ref{tab:timemulti} shows the antenna pattern averaged $\log{B_{SN}}$ for
five representative waveforms from the \texttt{RotCC} and \texttt{C\&S} models
injected in recolored noise. How well SMEE can distinguish a signal from noise is important 
because the explosion mechanism cannot be determined for a signal
it cannot detect. The table shows the mean $\log{B_{S,N}}$ is much larger
for waveforms from the \texttt{RotCC} model as they have a larger SNR than the 
\texttt{C\&S} waveforms. The value of $\log{B_{SN}}$ should be larger when using the PCs from the 
correct explosion mechanism. Waveforms from the \texttt{RotCC} model can be
distinguished from noise at all of the Galactic distances considered. Waveforms
from the \texttt{C\&S} model can all be distinguished from noise at $2\,$kpc.

\begin{table}[ht!]
\begin{center}
\begin{tabular}{|c|ccc|ccc|}
\hline
 Waveform & \multicolumn{3}{c|}{$\log{B_{\texttt{RottCC},N}}$}& \multicolumn{3}{c|}{$\log{B_{\texttt{C\&S},N}}$} \\
  & \multicolumn{1}{c}{2 kpc} & \multicolumn{1}{c}{10 kpc} & \multicolumn{1}{c|}{20 kpc}& \multicolumn{1}{c}{2 kpc}& \multicolumn{1}{c}{10 kpc} & \multicolumn{1}{c|}{20 kpc}\\  \hline \hline
\texttt{RotCC} & & & & & &\\
\phantom{0}s11a3o09\_shen & 24281 & 927 & 210 & 591 & 7 & -8 \\
\phantom{0}s15a2o09\_ls   & 27321 & 1050 & 241 & 785 & 15 & -7 \\
\phantom{0}s20a3o05\_ls   & 12151 & 447 & 92 & 1223 & 31 & -3 \\
\phantom{0}s40a3o07\_ls   & 54281 & 2121 & 508 & 1898 & 53 & 0  \\
\phantom{0}s40a3o13\_shen & 64323 & 2537 & 618 & 20510 & 815 & 192 \\  \hline \hline
\texttt{C\&S}  & & & & & &\\
\phantom{0}15\_3.2  & 52 & -4 & -5 & 328 & -6 & -12  \\
\phantom{0}15\_4.0  & 59 & -4 & -6 & 2982 & 90 & 5  \\
\phantom{0}20\_3.8  & 69 & -5 & -5 & 1629 & 352 & -8  \\
\phantom{0}40\_10.0 & 20 & -5 & -6 & 1687 & 42 & -4  \\
\phantom{0}40\_13.0 & 21 & -6 & -6 & 24 & -11 & -12  \\  \hline
\end{tabular}
\caption{The mean $\log{B_{S,N}}$ for five representative waveforms from each mechanism
injected at $2\,$kpc, $10\,$kpc and $20\,$kpc at the sky position of the
Galactic center. Waveforms from the \texttt{RottCC} model can be distinguished
from noise throughout the Galaxy. \texttt{C\&S} catalog waveforms at $20\,$kpc
are indistinguishable from noise.}
\label{tab:timemulti}
\end{center}
\end{table}

Fig.~\ref{fig:bfhists} shows histograms of $\log B_{\texttt{RotCC}-\texttt{C\&S}}$ for all
$1440$ injections at 3 Galactic distances. If the \texttt{RotCC} waveforms are identified with the
correct explosion mechanism then $\log B_{\texttt{RotCC}-\texttt{C\&S}}$ will be positive, and
if the \texttt{C\&S} waveforms are identified with the correct explosion
mechanism then $\log B_{\texttt{RotCC}-\texttt{C\&S}}$ will be negative. If $\log
B_{\texttt{RotCC}-\texttt{C\&S}}$ is between $-10$ and $10$ then either the injected waveform
could not be distinguished from noise or it is not possible to distinguish
between the explosion mechanisms considered.

The number of detected waveforms from the \texttt{C\&S} model is 157/160, 150/160 and 
19/160 at distances of $2\,$kpc, $10\,$kpc and $20\,$kpc respectively. The number of 
detected waveforms from the \texttt{RotCC} model is 1279/1280, 1198/1280 and 
1019/1280 at distances of $2\,$kpc, $10\,$kpc and $20\,$kpc respectively.
The correct explosion mechanism is determined for all detected waveforms from both models
at all Galactic distances.

All catalog waveforms are then injected at the sky position of the Large Magellanic Cloud at
a distance of $50\,$kpc at 10 different GPS times.
A histogram of $\log B_{\texttt{RotCC}-\texttt{C\&S}}$ is shown in
Fig.~\ref{fig:bfhists_d}. 707/1280 waveforms from the \texttt{RotCC}
model can be distinguished from noise at this distance and their explosion mechanism is correctly determined as
magnetorotational. Waveforms injected from the
\texttt{C\&S} model cannot be distinguished from noise at a distance of $50\,$kpc.

\subsection{Testing robustness using non-catalog waveforms}
As the waveforms from the \texttt{RotCC} and \texttt{C\&S} models used to create the PCs may not be
an exact match for a real CCSN GW signal it is important to test the robustness of
the method applied in SMEE using waveforms that do not come from the catalogs
used to construct the PCs. To test robustness we use five extra waveforms from each
mechanism. For the magnetorotational mechanism the extra waveforms are \texttt{sch1}, \texttt{sch2}, and 
the three \texttt{abd} waveforms as described in Section \ref{subsubsec:magrotGWcatalog}.
For the neutrino mechanism the five extra waveforms are the \texttt{yak}, \texttt{ott} and three 
\texttt{m\"uller} waveforms described in Section.~\ref{subsubsec:neutrinoGWcatalog}.

\begin{figure*}[ht!]
\begin{centering}
\subfigure[]{\label{fig:bfhists_extra_a}\includegraphics[width=8.5cm]{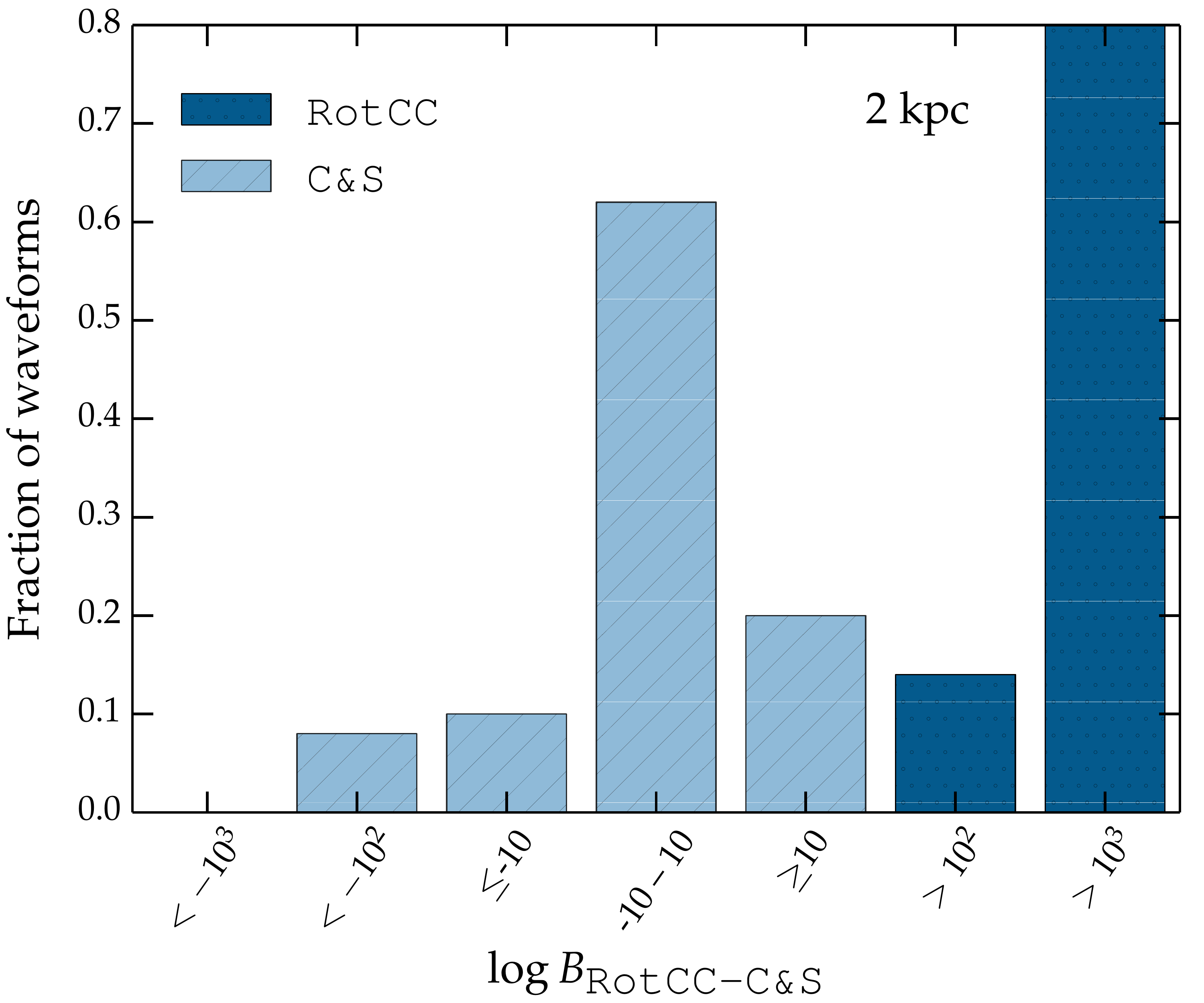}}
\subfigure[]{\label{fig:bfhists_extra_b}\includegraphics[width=8.5cm]{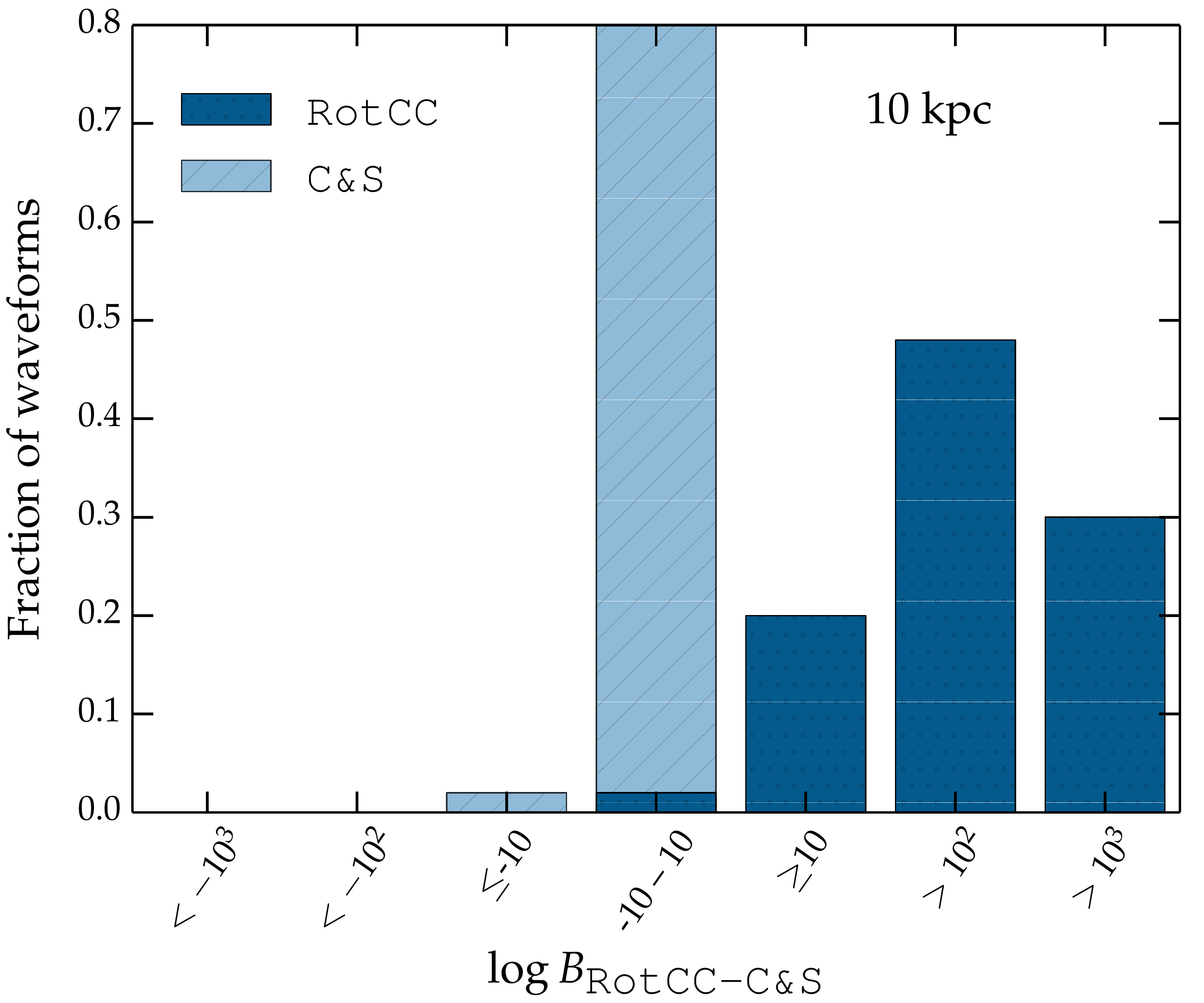}}
\subfigure[]{\label{fig:bfhists_extra_c}\includegraphics[width=8.5cm]{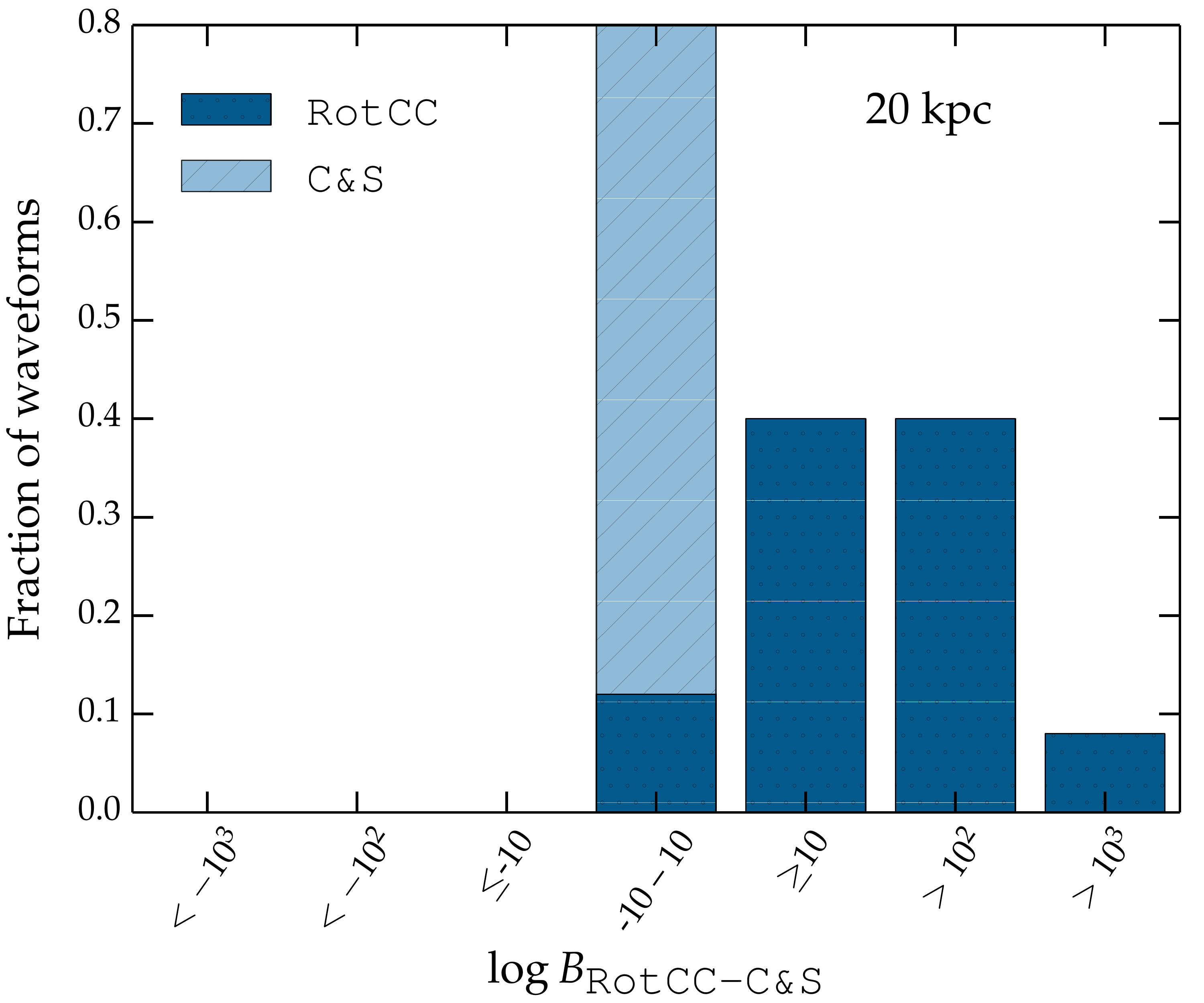}}
\subfigure[]{\label{fig:bfhists_extra_d}\includegraphics[width=8.5cm]{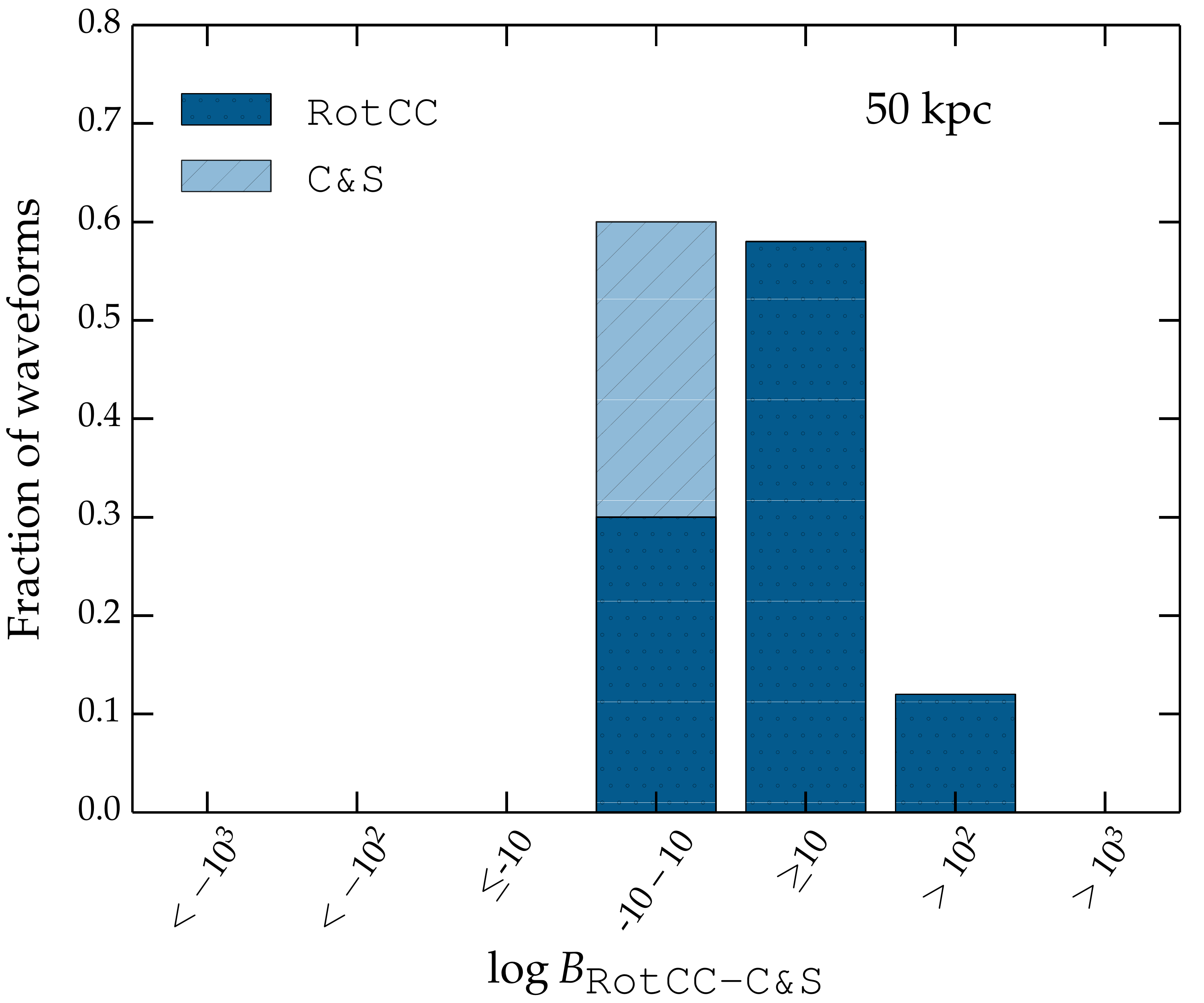}}
\caption{Log $B_{\texttt{RotCC}-\texttt{C\&S}}$ for 5 extra waveforms representing each explosion
mechanism. (a) At $2\,$kpc all extra magnetorotational mechanism waveforms can be distinguished 
from noise and their explosion mechanism is correctly determined. For the extra neutrino mechanism
waveforms only the explosion mechanism of the \texttt{yak} waveform is correctly determined. 
(b) At $10\,$kpc all extra neutrino mechanism waveforms cannot be distinguished from noise. (c) At 
$20\,$kpc 45/100 injected extra magnetorotational waveforms can be distinguished from noise and 
their explosion mechanism is correctly determined. (d) The correct explosion mechanism is determined 
for all extra magnetorotational waveforms distinguishable from noise (27/100) at $50\,$kpc.}
\label{fig:bfhists_extra}
\end{centering}
\end{figure*}
 
\begin{table}[h!]
\begin{center}
\begin{tabular}{|c|ccc|ccc|}
\hline
 Waveform & \multicolumn{3}{c|}{$\log{B_{\texttt{RotCC},N}}$}& \multicolumn{3}{c|}{$\log{B_{\texttt{C\&S},N}}$} \\
  & \multicolumn{1}{c}{2 kpc} & \multicolumn{1}{c}{10 kpc} & \multicolumn{1}{c|}{20 kpc}& \multicolumn{1}{c}{2 kpc}& \multicolumn{1}{c}{10 kpc} & \multicolumn{1}{c|}{20 kpc}\\  \hline \hline
 \texttt{RotCC}  & & & & & &\\
\phantom{0}\texttt{sch1} & 15116 & 567 & 124 & 2181 & 64 & 3 \\
\phantom{0}\texttt{sch2} & 47185 & 1843 & 441 & 7369 & 321 & 69 \\
\phantom{0}\texttt{abd1} & 87453 & 3454 & 843 & 21528 & 933 & 235 \\
\phantom{0}\texttt{abd2} & 50420 & 2000 & 488 & 18128 & 798 & 183 \\
\phantom{0}\texttt{abd3} & 6426 & 247 & 55 & 5147 & 185 & 31 \\  \hline \hline
\texttt{C\&S}  & & & & & &\\
\phantom{0}\texttt{yak}        & 23 & -5 & -6 & 141 & -10 & -11  \\
\phantom{0}\texttt{m\"uller1}  & -5 & -5 & -5 & -9 & -12 & -11  \\
\phantom{0}\texttt{m\"uller2}  & -5 & -6 & -5 & -8 & -10 & -12  \\
\phantom{0}\texttt{m\"uller3}  & -5 & -5 & -6 & -9 & -11 & -11 \\
\phantom{0}\texttt{ott}        & 118 & -2 & -6 & 24 & -12 & -12 \\  \hline
\end{tabular}
\caption{The mean $\log{B_{S,N}}$ for five extra waveforms representing each explosion mechanism
injected at $2\,$kpc, $10\,$kpc and $20\,$kpc at the sky position of the
Galactic center. The three \texttt{m\"uller} waveforms at $20\,$kpc are
indistinguishable from noise. The extra magnetorotational mechanism waveforms
can be distinguished from noise throughout our Galaxy. }
\label{tab:extratimemulti}
\end{center}
\end{table}

As for the \texttt{RotCC} and \texttt{C\&S} waveforms the 10 extra waveforms are injected at 10 GPS times at
the sky position of the Galactic center at distances of $2\,$kpc, $10\,$kpc and
$20\,$kpc leading to a total of 100 injections at each distance. Table.~\ref{tab:extratimemulti} shows how well 
the extra waveforms can be distinguished from noise at the three Galactic distances considered. As for the
catalog waveforms the table shows the antenna pattern averaged values of
$\log{B_{S,N}}$. A larger value of $\log{B_{S,N}}$ is expected when the correct PCs are used. The confidence in the result is larger for larger values of $\log{B_{S,N}}$. All the extra magnetorotational mechanism waveforms can be
distinguished from noise at the 3 Galactic distances considered. The \texttt{yak} and
\texttt{ott} waveforms can be distinguished from noise at $2\,$kpc.
The three \texttt{m\"uller} waveforms cannot be
distinguished from noise at any of the Galactic distances considered.

Fig.~\ref{fig:bfhists_extra} shows histograms of $\log B_{\texttt{RotCC}-\texttt{C\&S}}$ for all
100 extra waveform injections at distances throughout the Galaxy.
As for the waveforms used to calculate the PCs if the explosion mechanism of the magnetorotational waveforms are
correctly determined then $\log B_{\texttt{RotCC}-\texttt{C\&S}}$ will be positive and if the explosion
mechanism of the neutrino mechanism waveforms are correctly determined 
then $\log B_{\texttt{RotCC}-\texttt{C\&S}}$ will be negative. 
At all distances the 30 injected \texttt{m\"uller} waveforms cannot be distinguished from noise.
At $2\,$kpc the explosion mechanism of the 10 injected \texttt{yak} waveforms is correctly determined as 
neutrino-driven. 
The explosion mechanism of the 10 \texttt{ott} waveform injections are incorrectly determined as
magnetorotational. 
The \texttt{Ott} waveforms, shown in Fig. \ref{fig:waveforms_3d}, contain a feature
during the first $20\,$ms that appears reminiscent of the rotational bounce signals.
This is due to a strong signal from the early post-bounce phase that arises because of
artificially strong prompt convection induced by the neutrino leakage scheme. This feature is 
likely the cause of the incorrect result.
If larger catalogs of 3D CCSN waveforms are obtained then PCs 
containing both polarizations could be used to improve results for any waveforms that are
currently poorly reconstructed by SMEE.
All extra magnetorotational mechanism injections at $2\,$kpc are distinguished from noise
and their explosion mechanism is correctly determined.

At $10\,$kpc 1/10 \texttt{yak} injections and 49/50 magnetorotational injected waveforms can be distinguished 
from noise. The explosion mechanism is correctly determined for all detected waveforms.   
At $20\,$kpc 45/50 magnetorotational waveforms and none of the extra neutrino mechanism waveforms can be distinguished 
from noise. The explosion mechanism is correctly determined for all detected
magnetorotational waveforms at $20\,$kpc.

Fig.~\ref{fig:bfhists_extra_d} shows a histogram of $\log B_{\texttt{RotCC}-\texttt{C\&S}}$ for
100 injections of the extra waveforms at $50\,$kpc at the sky position of the Large
Magellanic Cloud. 27/50 magnetorotational waveforms can be distinguished from noise and their
explosion mechanism is correctly determined as magnetorotational.


\section{SUMMARY AND DISCUSSION}
\label{sec:discussion}

The Supernova Model Evidence Extractor (SMEE) is designed to measure astrophysical
parameters of a CCSN GW detection.
CCSNe have long been considered as a potential source for 
an aLIGO and AdVirgo detector network and a CCSN detection may provide an ideal
probe of the inner regions of the explosion that do not emit
electromagnetically. Determining the CCSN explosion mechanism is essential 
for a full understanding of the physics and processes involved in CCSNe. 

For the first time we demonstrate the ability of SMEE to determine the CCSN
explosion mechanism with a network of GW detectors with real 
non-stationary and non-Gaussian noise. In this paper SMEE considers the 
magnetorotational and neutrino explosion mechanisms and shows how 
the correct explosion mechanism can be determined for all detectable catalog waveforms at
distances throughout our Galaxy. GW signals from neutrino-driven convection have
a smaller amplitude than those from rapidly-rotating core collapse, and therefore
detections at distances of $10\,$kpc or less are needed for a robust result. Furthermore, we
can determine the explosion mechanism of rapidly-rotating core collapse
waveforms at the distance and sky position of the Large Magellanic Cloud.

We further enhance the model
selection capabilities of SMEE with a careful selection of the number of PCs that
considers the relative complexity of the different explosion models.
A large number of PCs is required to represent all the common features of the
neutrino-driven convection waveforms. The number of available waveforms is much
smaller than those available for rapidly-rotating core collapse and the differences between 
individual waveforms is much larger. This leads to a reduction in the robustness of the
result from SMEE as the parameter space of the neutrino waveforms is not sufficiently covered.
Furthermore, 3D neutrino waveforms contain some features that are different from the 2D 
waveforms used to create the PCs. However, the 2D rapidly-rotating core collapse waveforms are
still a good approximation for 3D rapidly-rotating waveforms as nonaxisymmetric
instabilities occur after the signal bounce that is the main feature in the rapidly-rotating PCs. 

During recent years 2D neutrino mechanism waveforms with more detailed physics have 
become available. They include an updated version of the 
\texttt{yak} waveforms used in this study, which are now complete (up to 1s) waveforms, as 
the 2010 waveforms were truncated at $\sim500\,$ms after bounce \cite{yakunin:15}.
Waveforms produced by M\"uller~\emph{et al.} (2013) \cite{muller:13} are the first 2D CCSNe 
relativistic GW signals with multi-group, three-flavour neutrino transport.  
Furthermore, a larger number of 3D neutrino mechanism waveforms have 
become available recently, including Kuroda~\emph{et al.} (2016) \cite{kuroda:16} who simulate a $15M_{\odot}$ 
star with three different EOSs showing a strong low-frequency signal from the SASI, and 
Andresen~\emph{et al.} (2016) \cite{andresen:16} that include multi-group neutrino transport.
Updating SMEE to use these 3D waveforms, as well as other 3D waveforms for rapidly-rotating
CCSNe (e.g. Kuroda~\emph{et al.} \cite{kuroda:14}), will be essential for future robust 
parameter estimation with CCSNe GWs.   

Future work for SMEE will include following up real GW triggers found in the searches for 
GWs as it is possible that a real trigger may not belong to any of 
the models considered by SMEE. 
Therefore, future work for SMEE will include distinguishing an astrophysical CCSN
explosion mechanism signal from other GW signal types and noise transients. 
Spectrograms or power spectra may be used instead of the Fourier transform of the 
time series waveforms to remove the models reliance on phase. How well SMEE can 
reconstruct the detected GW signal will also be explored in future studies. This 
can be compared with other tools that reconstruct GW waveforms using minimal assumptions
about the signal morphology \cite{bayeswave, 2016PhRvD..93d2004K, 0264-9381-25-11-114029}.


\begin{acknowledgments}
The authors acknowledge helpful exchanges with James Clark, Alan Weinstein,
Jonah Kanner, Rory Smith, and the LIGO SN working group that have benefitted this 
paper. We thank the CCSN simulation community for making
their gravitational waveform predictions available for this
study. ISH, JP, and JL are supported by UK 
Science and Technology Facilities Council grants ST/L000946/1 and ST/L000946/1.
The authors also gratefully acknowledge the support of the Scottish 
Universities Physics Alliance (SUPA). LIGO was constructed by the California Institute of Technology
and Massachusetts Institute of Technology with funding from the
National Science Foundation and operates under cooperative agreements
PHY-0107417 and PHY-0757058. This paper has been assigned LIGO Document
Number LIGO-P1600289.
\end{acknowledgments}


\bibliographystyle{apsrev}

\bibliography{smee_bib}

\begin{thebibliography}{87}
\expandafter\ifx\csname natexlab\endcsname\relax\def\natexlab#1{#1}\fi
\expandafter\ifx\csname bibnamefont\endcsname\relax
  \def\bibnamefont#1{#1}\fi
\expandafter\ifx\csname bibfnamefont\endcsname\relax
  \def\bibfnamefont#1{#1}\fi
\expandafter\ifx\csname citenamefont\endcsname\relax
  \def\citenamefont#1{#1}\fi
\expandafter\ifx\csname url\endcsname\relax
  \def\url#1{\texttt{#1}}\fi
\expandafter\ifx\csname urlprefix\endcsname\relax\def\urlprefix{URL }\fi
\providecommand{\bibinfo}[2]{#2}
\providecommand{\eprint}[2][]{\url{#2}}

\bibitem[{\citenamefont{{Baade} and {Zwicky}}(1934)}]{baade:34}
\bibinfo{author}{\bibfnamefont{W.}~\bibnamefont{{Baade}}} \bibnamefont{and}
  \bibinfo{author}{\bibfnamefont{F.}~\bibnamefont{{Zwicky}}},
  \bibinfo{journal}{Proceedings of the National Academy of Science}
  \textbf{\bibinfo{volume}{20}}, \bibinfo{pages}{259} (\bibinfo{year}{1934}).

\bibitem[{\citenamefont{{Baron} and {Cooperstein}}(1990)}]{baron:90}
\bibinfo{author}{\bibfnamefont{E.}~\bibnamefont{{Baron}}} \bibnamefont{and}
  \bibinfo{author}{\bibfnamefont{J.}~\bibnamefont{{Cooperstein}}},
  \bibinfo{journal}{\apj} \textbf{\bibinfo{volume}{353}}, \bibinfo{pages}{597}
  (\bibinfo{year}{1990}).

\bibitem[{\citenamefont{{Bethe}}(1990)}]{bethe:90}
\bibinfo{author}{\bibfnamefont{H.~A.} \bibnamefont{{Bethe}}},
  \bibinfo{journal}{Reviews of Modern Physics} \textbf{\bibinfo{volume}{62}},
  \bibinfo{pages}{801} (\bibinfo{year}{1990}).

\bibitem[{\citenamefont{{O'Connor} and {Ott}}(2011)}]{oconnor:11}
\bibinfo{author}{\bibfnamefont{E.}~\bibnamefont{{O'Connor}}} \bibnamefont{and}
  \bibinfo{author}{\bibfnamefont{C.~D.} \bibnamefont{{Ott}}},
  \bibinfo{journal}{\apj} \textbf{\bibinfo{volume}{730}}, \bibinfo{eid}{70}
  (\bibinfo{year}{2011}).

\bibitem[{\citenamefont{{Janka}}(2012)}]{janka:12a}
\bibinfo{author}{\bibfnamefont{H.-T.} \bibnamefont{{Janka}}},
  \bibinfo{journal}{Ann. Rev. Nuc. Par. Sci.} \textbf{\bibinfo{volume}{62}},
  \bibinfo{pages}{407} (\bibinfo{year}{2012}).

\bibitem[{\citenamefont{{Foglizzo} et~al.}(2015)\citenamefont{{Foglizzo},
  {Kazeroni}, {Guilet}, {Masset}, {Gonz{\'a}lez}, {Krueger}, {Novak}, {Oertel},
  {Margueron}, {Faure} et~al.}}]{foglizzo:15}
\bibinfo{author}{\bibfnamefont{T.}~\bibnamefont{{Foglizzo}}},
  \bibinfo{author}{\bibfnamefont{R.}~\bibnamefont{{Kazeroni}}},
  \bibinfo{author}{\bibfnamefont{J.}~\bibnamefont{{Guilet}}},
  \bibinfo{author}{\bibfnamefont{F.}~\bibnamefont{{Masset}}},
  \bibinfo{author}{\bibfnamefont{M.}~\bibnamefont{{Gonz{\'a}lez}}},
  \bibinfo{author}{\bibfnamefont{B.~K.} \bibnamefont{{Krueger}}},
  \bibinfo{author}{\bibfnamefont{J.}~\bibnamefont{{Novak}}},
  \bibinfo{author}{\bibfnamefont{M.}~\bibnamefont{{Oertel}}},
  \bibinfo{author}{\bibfnamefont{J.}~\bibnamefont{{Margueron}}},
  \bibinfo{author}{\bibfnamefont{J.}~\bibnamefont{{Faure}}},
  \bibnamefont{et~al.}, \bibinfo{journal}{pasa} \textbf{\bibinfo{volume}{32}},
  \bibinfo{eid}{e009} (\bibinfo{year}{2015}).

\bibitem[{\citenamefont{{Hirata} et~al.}(1987)\citenamefont{{Hirata}, {Kajita},
  {Koshiba}, {Nakahata}, and {Oyama}}}]{hirata:87}
\bibinfo{author}{\bibfnamefont{K.}~\bibnamefont{{Hirata}}},
  \bibinfo{author}{\bibfnamefont{T.}~\bibnamefont{{Kajita}}},
  \bibinfo{author}{\bibfnamefont{M.}~\bibnamefont{{Koshiba}}},
  \bibinfo{author}{\bibfnamefont{M.}~\bibnamefont{{Nakahata}}},
  \bibnamefont{and} \bibinfo{author}{\bibfnamefont{Y.}~\bibnamefont{{Oyama}}},
  \bibinfo{journal}{Physical Review Letters} \textbf{\bibinfo{volume}{58}},
  \bibinfo{pages}{1490} (\bibinfo{year}{1987}).

\bibitem[{\citenamefont{{Bionta} et~al.}(1987)\citenamefont{{Bionta},
  {Blewitt}, {Bratton}, {Casper}, and {Ciocio}}}]{bionta:87}
\bibinfo{author}{\bibfnamefont{R.~M.} \bibnamefont{{Bionta}}},
  \bibinfo{author}{\bibfnamefont{G.}~\bibnamefont{{Blewitt}}},
  \bibinfo{author}{\bibfnamefont{C.~B.} \bibnamefont{{Bratton}}},
  \bibinfo{author}{\bibfnamefont{D.}~\bibnamefont{{Casper}}}, \bibnamefont{and}
  \bibinfo{author}{\bibfnamefont{A.}~\bibnamefont{{Ciocio}}},
  \bibinfo{journal}{Physical Review Letters} \textbf{\bibinfo{volume}{58}},
  \bibinfo{pages}{1494} (\bibinfo{year}{1987}).

\bibitem[{\citenamefont{{Schaeffer} et~al.}(1987)\citenamefont{{Schaeffer},
  {Declais}, and {Jullian}}}]{schaeffer:87}
\bibinfo{author}{\bibfnamefont{R.}~\bibnamefont{{Schaeffer}}},
  \bibinfo{author}{\bibfnamefont{Y.}~\bibnamefont{{Declais}}},
  \bibnamefont{and}
  \bibinfo{author}{\bibfnamefont{S.}~\bibnamefont{{Jullian}}},
  \bibinfo{journal}{\nat} \textbf{\bibinfo{volume}{330}}, \bibinfo{pages}{142}
  (\bibinfo{year}{1987}).

\bibitem[{\citenamefont{{Ott}}(2009)}]{ott:09}
\bibinfo{author}{\bibfnamefont{C.~D.} \bibnamefont{{Ott}}},
  \bibinfo{journal}{Class. Quantum Grav.} \textbf{\bibinfo{volume}{26}},
  \bibinfo{pages}{063001} (\bibinfo{year}{2009}).

\bibitem[{\citenamefont{{Kotake} et~al.}(2006)\citenamefont{{Kotake}, {Sato},
  and {Takahashi}}}]{kotake:06}
\bibinfo{author}{\bibfnamefont{K.}~\bibnamefont{{Kotake}}},
  \bibinfo{author}{\bibfnamefont{K.}~\bibnamefont{{Sato}}}, \bibnamefont{and}
  \bibinfo{author}{\bibfnamefont{K.}~\bibnamefont{{Takahashi}}},
  \bibinfo{journal}{Reports on Progress in Physics}
  \textbf{\bibinfo{volume}{69}}, \bibinfo{pages}{971} (\bibinfo{year}{2006}).

\bibitem[{\citenamefont{{The LIGO Scientific Collaboration}
  et~al.}(2015)\citenamefont{{The LIGO Scientific Collaboration}, {Aasi},
  {Abbott}, {Abbott}, and et~al.}}]{aLIGO}
\bibinfo{author}{\bibnamefont{{The LIGO Scientific Collaboration}}},
  \bibinfo{author}{\bibfnamefont{J.}~\bibnamefont{{Aasi}}},
  \bibinfo{author}{\bibfnamefont{B.~P.} \bibnamefont{{Abbott}}},
  \bibinfo{author}{\bibfnamefont{R.}~\bibnamefont{{Abbott}}}, \bibnamefont{and}
  \bibinfo{author}{\bibnamefont{et~al.}}, \bibinfo{journal}{\cqg}
  \textbf{\bibinfo{volume}{32}}, \bibinfo{eid}{074001} (\bibinfo{year}{2015}).

\bibitem[{\citenamefont{{Acernese} and et~al.}(2015)}]{AdVirgo}
\bibinfo{author}{\bibfnamefont{F.}~\bibnamefont{{Acernese}}} \bibnamefont{and}
  \bibinfo{author}{\bibnamefont{et~al.}}, \bibinfo{journal}{\cqg}
  \textbf{\bibinfo{volume}{32}}, \bibinfo{eid}{024001} (\bibinfo{year}{2015}).

\bibitem[{\citenamefont{{Gossan} et~al.}(2016)\citenamefont{{Gossan}, {Sutton},
  {Stuver}, {Zanolin}, {Gill}, and {Ott}}}]{gossan:16}
\bibinfo{author}{\bibfnamefont{S.~E.} \bibnamefont{{Gossan}}},
  \bibinfo{author}{\bibfnamefont{P.}~\bibnamefont{{Sutton}}},
  \bibinfo{author}{\bibfnamefont{A.}~\bibnamefont{{Stuver}}},
  \bibinfo{author}{\bibfnamefont{M.}~\bibnamefont{{Zanolin}}},
  \bibinfo{author}{\bibfnamefont{K.}~\bibnamefont{{Gill}}}, \bibnamefont{and}
  \bibinfo{author}{\bibfnamefont{C.~D.} \bibnamefont{{Ott}}},
  \bibinfo{journal}{\prd} \textbf{\bibinfo{volume}{93}}, \bibinfo{eid}{042002}
  (\bibinfo{year}{2016}).

\bibitem[{\citenamefont{{van den Bergh} and
  {Tammann}}(1991)}]{1991ARA&A..29..363V}
\bibinfo{author}{\bibfnamefont{S.}~\bibnamefont{{van den Bergh}}}
  \bibnamefont{and} \bibinfo{author}{\bibfnamefont{G.~A.}
  \bibnamefont{{Tammann}}}, \bibinfo{journal}{\araa}
  \textbf{\bibinfo{volume}{29}}, \bibinfo{pages}{363} (\bibinfo{year}{1991}).

\bibitem[{\citenamefont{{Cappellaro} et~al.}(1993)\citenamefont{{Cappellaro},
  {Turatto}, {Benetti}, {Tsvetkov}, {Bartunov}, and
  {Makarova}}}]{1993A&A...273..383C}
\bibinfo{author}{\bibfnamefont{E.}~\bibnamefont{{Cappellaro}}},
  \bibinfo{author}{\bibfnamefont{M.}~\bibnamefont{{Turatto}}},
  \bibinfo{author}{\bibfnamefont{S.}~\bibnamefont{{Benetti}}},
  \bibinfo{author}{\bibfnamefont{D.~Y.} \bibnamefont{{Tsvetkov}}},
  \bibinfo{author}{\bibfnamefont{O.~S.} \bibnamefont{{Bartunov}}},
  \bibnamefont{and} \bibinfo{author}{\bibfnamefont{I.~N.}
  \bibnamefont{{Makarova}}}, \bibinfo{journal}{\aap}
  \textbf{\bibinfo{volume}{273}}, \bibinfo{pages}{383} (\bibinfo{year}{1993}).

\bibitem[{\citenamefont{Alexeyev and Alexeyeva}(2002)}]{Alexeyev2002}
\bibinfo{author}{\bibfnamefont{E.~N.} \bibnamefont{Alexeyev}} \bibnamefont{and}
  \bibinfo{author}{\bibfnamefont{L.~N.} \bibnamefont{Alexeyeva}},
  \bibinfo{journal}{Journal of Experimental and Theoretical Physics}
  \textbf{\bibinfo{volume}{95}}, \bibinfo{pages}{5} (\bibinfo{year}{2002}),
  ISSN \bibinfo{issn}{1090-6509},
  \urlprefix\url{http://dx.doi.org/10.1134/1.1499896}.

\bibitem[{\citenamefont{Gill et~al.}(2016)\citenamefont{Gill, Branchesi,
  Zanolin, and M.}}]{rates:jaz}
\bibinfo{author}{\bibfnamefont{K.}~\bibnamefont{Gill}},
  \bibinfo{author}{\bibfnamefont{M.}~\bibnamefont{Branchesi}},
  \bibinfo{author}{\bibfnamefont{M.}~\bibnamefont{Zanolin}}, \bibnamefont{and}
  \bibinfo{author}{\bibfnamefont{S.}~\bibnamefont{M.}}, \bibinfo{journal}{In
  prep}  (\bibinfo{year}{2016}).

\bibitem[{\citenamefont{{Abbott} et~al.}(2016)}]{2016arXiv160501785A}
\bibinfo{author}{\bibfnamefont{B.~P.} \bibnamefont{{Abbott}}}
  \bibnamefont{et~al.}, \bibinfo{journal}{ArXiv e-prints}
  (\bibinfo{year}{2016}).

\bibitem[{\citenamefont{{Owen} and {Sathyaprakash}}(1999)}]{owen:99}
\bibinfo{author}{\bibfnamefont{B.~J.} \bibnamefont{{Owen}}} \bibnamefont{and}
  \bibinfo{author}{\bibfnamefont{B.~S.} \bibnamefont{{Sathyaprakash}}},
  \bibinfo{journal}{\prd} \textbf{\bibinfo{volume}{60}}, \bibinfo{eid}{022002}
  (\bibinfo{year}{1999}).

\bibitem[{\citenamefont{{Logue} et~al.}(2012)\citenamefont{{Logue}, {Ott},
  {Heng}, {Kalmus}, and {Scargill}}}]{logue:12}
\bibinfo{author}{\bibfnamefont{J.}~\bibnamefont{{Logue}}},
  \bibinfo{author}{\bibfnamefont{C.~D.} \bibnamefont{{Ott}}},
  \bibinfo{author}{\bibfnamefont{I.~S.} \bibnamefont{{Heng}}},
  \bibinfo{author}{\bibfnamefont{P.}~\bibnamefont{{Kalmus}}}, \bibnamefont{and}
  \bibinfo{author}{\bibfnamefont{J.}~\bibnamefont{{Scargill}}},
  \bibinfo{journal}{\prd} \textbf{\bibinfo{volume}{86}}, \bibinfo{eid}{044023}
  (\bibinfo{year}{2012}).

\bibitem[{\citenamefont{{Janka} et~al.}(2007)\citenamefont{{Janka}, {Langanke},
  {Marek}, {Mart{\'{\i}}nez-Pinedo}, and {M{\"u}ller}}}]{janka:07}
\bibinfo{author}{\bibfnamefont{H.-T.} \bibnamefont{{Janka}}},
  \bibinfo{author}{\bibfnamefont{K.}~\bibnamefont{{Langanke}}},
  \bibinfo{author}{\bibfnamefont{A.}~\bibnamefont{{Marek}}},
  \bibinfo{author}{\bibfnamefont{G.}~\bibnamefont{{Mart{\'{\i}}nez-Pinedo}}},
  \bibnamefont{and}
  \bibinfo{author}{\bibfnamefont{B.}~\bibnamefont{{M{\"u}ller}}},
  \bibinfo{journal}{\physrep} \textbf{\bibinfo{volume}{442}},
  \bibinfo{pages}{38} (\bibinfo{year}{2007}).

\bibitem[{\citenamefont{{Burrows}
  et~al.}(2007{\natexlab{a}})\citenamefont{{Burrows}, {Dessart}, {Livne},
  {Ott}, and {Murphy}}}]{burrows:07b}
\bibinfo{author}{\bibfnamefont{A.}~\bibnamefont{{Burrows}}},
  \bibinfo{author}{\bibfnamefont{L.}~\bibnamefont{{Dessart}}},
  \bibinfo{author}{\bibfnamefont{E.}~\bibnamefont{{Livne}}},
  \bibinfo{author}{\bibfnamefont{C.~D.} \bibnamefont{{Ott}}}, \bibnamefont{and}
  \bibinfo{author}{\bibfnamefont{J.}~\bibnamefont{{Murphy}}},
  \bibinfo{journal}{\apj} \textbf{\bibinfo{volume}{664}}, \bibinfo{pages}{416}
  (\bibinfo{year}{2007}{\natexlab{a}}).

\bibitem[{\citenamefont{{Burrows} et~al.}(2006)\citenamefont{{Burrows},
  {Livne}, {Dessart}, {Ott}, and {Murphy}}}]{burrows:06}
\bibinfo{author}{\bibfnamefont{A.}~\bibnamefont{{Burrows}}},
  \bibinfo{author}{\bibfnamefont{E.}~\bibnamefont{{Livne}}},
  \bibinfo{author}{\bibfnamefont{L.}~\bibnamefont{{Dessart}}},
  \bibinfo{author}{\bibfnamefont{C.~D.} \bibnamefont{{Ott}}}, \bibnamefont{and}
  \bibinfo{author}{\bibfnamefont{J.}~\bibnamefont{{Murphy}}},
  \bibinfo{journal}{\apj} \textbf{\bibinfo{volume}{640}}, \bibinfo{pages}{878}
  (\bibinfo{year}{2006}).

\bibitem[{\citenamefont{{Burrows}
  et~al.}(2007{\natexlab{b}})\citenamefont{{Burrows}, {Livne}, {Dessart},
  {Ott}, and {Murphy}}}]{burrows:07a}
\bibinfo{author}{\bibfnamefont{A.}~\bibnamefont{{Burrows}}},
  \bibinfo{author}{\bibfnamefont{E.}~\bibnamefont{{Livne}}},
  \bibinfo{author}{\bibfnamefont{L.}~\bibnamefont{{Dessart}}},
  \bibinfo{author}{\bibfnamefont{C.~D.} \bibnamefont{{Ott}}}, \bibnamefont{and}
  \bibinfo{author}{\bibfnamefont{J.}~\bibnamefont{{Murphy}}},
  \bibinfo{journal}{\apj} \textbf{\bibinfo{volume}{655}}, \bibinfo{pages}{416}
  (\bibinfo{year}{2007}{\natexlab{b}}).

\bibitem[{\citenamefont{{Heng}}(2009)}]{heng:09}
\bibinfo{author}{\bibfnamefont{I.~S.} \bibnamefont{{Heng}}},
  \bibinfo{journal}{\cqg} \textbf{\bibinfo{volume}{26}}, \bibinfo{eid}{105005}
  (\bibinfo{year}{2009}).

\bibitem[{\citenamefont{{R{\"o}ver} et~al.}(2009)\citenamefont{{R{\"o}ver},
  {Bizouard}, {Christensen}, {Dimmelmeier}, {Heng}, and {Meyer}}}]{roever:09}
\bibinfo{author}{\bibfnamefont{C.}~\bibnamefont{{R{\"o}ver}}},
  \bibinfo{author}{\bibfnamefont{M.-A.} \bibnamefont{{Bizouard}}},
  \bibinfo{author}{\bibfnamefont{N.}~\bibnamefont{{Christensen}}},
  \bibinfo{author}{\bibfnamefont{H.}~\bibnamefont{{Dimmelmeier}}},
  \bibinfo{author}{\bibfnamefont{I.~S.} \bibnamefont{{Heng}}},
  \bibnamefont{and} \bibinfo{author}{\bibfnamefont{R.}~\bibnamefont{{Meyer}}},
  \bibinfo{journal}{\prd} \textbf{\bibinfo{volume}{80}}, \bibinfo{eid}{102004}
  (\bibinfo{year}{2009}).

\bibitem[{\citenamefont{{Mardia} et~al.}(1979)\citenamefont{{Mardia}, {Kent},
  and {Bibby}}}]{mardia:79}
\bibinfo{author}{\bibfnamefont{K.~V.} \bibnamefont{{Mardia}}},
  \bibinfo{author}{\bibfnamefont{J.~T.} \bibnamefont{{Kent}}},
  \bibnamefont{and} \bibinfo{author}{\bibfnamefont{J.~M.}
  \bibnamefont{{Bibby}}}, \emph{\bibinfo{title}{{Multivariate analysis}}}
  (\bibinfo{year}{1979}).

\bibitem[{\citenamefont{{Shoemaker}}(2010)}]{aLIGOdesign}
\bibinfo{author}{\bibfnamefont{D.}~\bibnamefont{{Shoemaker}}},
  \bibinfo{type}{Tech. Rep.} (\bibinfo{year}{2010}).

\bibitem[{\citenamefont{{Skilling}}(2004)}]{skilling:04}
\bibinfo{author}{\bibfnamefont{J.}~\bibnamefont{{Skilling}}}, in
  \emph{\bibinfo{booktitle}{American Institute of Physics Conference Series}},
  edited by \bibinfo{editor}{\bibfnamefont{R.}~\bibnamefont{{Fischer}}},
  \bibinfo{editor}{\bibfnamefont{R.}~\bibnamefont{{Preuss}}}, \bibnamefont{and}
  \bibinfo{editor}{\bibfnamefont{U.~V.} \bibnamefont{{Toussaint}}}
  (\bibinfo{year}{2004}), vol. \bibinfo{volume}{735} of
  \emph{\bibinfo{series}{American Institute of Physics Conference Series}}, pp.
  \bibinfo{pages}{395--405}.

\bibitem[{\citenamefont{{Wang} and {Wheeler}}(2008)}]{wang:08}
\bibinfo{author}{\bibfnamefont{L.}~\bibnamefont{{Wang}}} \bibnamefont{and}
  \bibinfo{author}{\bibfnamefont{J.~C.} \bibnamefont{{Wheeler}}},
  \bibinfo{journal}{\araa} \textbf{\bibinfo{volume}{46}}, \bibinfo{pages}{433}
  (\bibinfo{year}{2008}).

\bibitem[{\citenamefont{{Chornock} et~al.}(2011)}]{chornock:11}
\bibinfo{author}{\bibfnamefont{R.}~\bibnamefont{{Chornock}}}
  \bibnamefont{et~al.}, \bibinfo{journal}{\apj} \textbf{\bibinfo{volume}{739}},
  \bibinfo{eid}{41} (\bibinfo{year}{2011}).

\bibitem[{\citenamefont{{Smith} et~al.}(2012)}]{smith:12a}
\bibinfo{author}{\bibfnamefont{N.}~\bibnamefont{{Smith}}} \bibnamefont{et~al.},
  \bibinfo{journal}{\mnras} \textbf{\bibinfo{volume}{420}},
  \bibinfo{pages}{1135} (\bibinfo{year}{2012}).

\bibitem[{\citenamefont{{Sinnott} et~al.}(2013)\citenamefont{{Sinnott},
  {Welch}, {Rest}, {Sutherland}, and {Bergmann}}}]{sinnott:13}
\bibinfo{author}{\bibfnamefont{B.}~\bibnamefont{{Sinnott}}},
  \bibinfo{author}{\bibfnamefont{D.~L.} \bibnamefont{{Welch}}},
  \bibinfo{author}{\bibfnamefont{A.}~\bibnamefont{{Rest}}},
  \bibinfo{author}{\bibfnamefont{P.~G.} \bibnamefont{{Sutherland}}},
  \bibnamefont{and}
  \bibinfo{author}{\bibfnamefont{M.}~\bibnamefont{{Bergmann}}},
  \bibinfo{journal}{\apj} \textbf{\bibinfo{volume}{767}}, \bibinfo{eid}{45}
  (\bibinfo{year}{2013}).

\bibitem[{\citenamefont{{Boggs} et~al.}(2015)}]{boggs:15}
\bibinfo{author}{\bibfnamefont{S.~E.} \bibnamefont{{Boggs}}}
  \bibnamefont{et~al.}, \bibinfo{journal}{Science}
  \textbf{\bibinfo{volume}{348}}, \bibinfo{pages}{670} (\bibinfo{year}{2015}).

\bibitem[{\citenamefont{{Ott} et~al.}(2013)}]{ott:13a}
\bibinfo{author}{\bibfnamefont{C.~D.} \bibnamefont{{Ott}}}
  \bibnamefont{et~al.}, \bibinfo{journal}{\apj} \textbf{\bibinfo{volume}{768}},
  \bibinfo{eid}{115} (\bibinfo{year}{2013}).

\bibitem[{\citenamefont{{Kuroda} et~al.}(2016)\citenamefont{{Kuroda}, {Kotake},
  and {Takiwaki}}}]{kuroda:16}
\bibinfo{author}{\bibfnamefont{T.}~\bibnamefont{{Kuroda}}},
  \bibinfo{author}{\bibfnamefont{K.}~\bibnamefont{{Kotake}}}, \bibnamefont{and}
  \bibinfo{author}{\bibfnamefont{T.}~\bibnamefont{{Takiwaki}}},
  \bibinfo{journal}{\apjl} \textbf{\bibinfo{volume}{829}}, \bibinfo{eid}{L14}
  (\bibinfo{year}{2016}).

\bibitem[{\citenamefont{{Ott} et~al.}(2007)\citenamefont{{Ott}, {Dimmelmeier},
  {Marek}, {Janka}, {Hawke}, {Zink}, and {Schnetter}}}]{ott:07}
\bibinfo{author}{\bibfnamefont{C.~D.} \bibnamefont{{Ott}}},
  \bibinfo{author}{\bibfnamefont{H.}~\bibnamefont{{Dimmelmeier}}},
  \bibinfo{author}{\bibfnamefont{A.}~\bibnamefont{{Marek}}},
  \bibinfo{author}{\bibfnamefont{H.-T.} \bibnamefont{{Janka}}},
  \bibinfo{author}{\bibfnamefont{I.}~\bibnamefont{{Hawke}}},
  \bibinfo{author}{\bibfnamefont{B.}~\bibnamefont{{Zink}}}, \bibnamefont{and}
  \bibinfo{author}{\bibfnamefont{E.}~\bibnamefont{{Schnetter}}},
  \bibinfo{journal}{Physical Review Letters} \textbf{\bibinfo{volume}{98}},
  \bibinfo{eid}{261101} (\bibinfo{year}{2007}).

\bibitem[{\citenamefont{{Shibata} and {Sekiguchi}}(2005)}]{shibata:05}
\bibinfo{author}{\bibfnamefont{M.}~\bibnamefont{{Shibata}}} \bibnamefont{and}
  \bibinfo{author}{\bibfnamefont{Y.-I.} \bibnamefont{{Sekiguchi}}},
  \bibinfo{journal}{\prd} \textbf{\bibinfo{volume}{71}}, \bibinfo{eid}{024014}
  (\bibinfo{year}{2005}).

\bibitem[{\citenamefont{{Scheidegger} et~al.}(2008)\citenamefont{{Scheidegger},
  {Fischer}, {Whitehouse}, and {Liebend{\"o}rfer}}}]{scheidegger:08}
\bibinfo{author}{\bibfnamefont{S.}~\bibnamefont{{Scheidegger}}},
  \bibinfo{author}{\bibfnamefont{T.}~\bibnamefont{{Fischer}}},
  \bibinfo{author}{\bibfnamefont{S.~C.} \bibnamefont{{Whitehouse}}},
  \bibnamefont{and}
  \bibinfo{author}{\bibfnamefont{M.}~\bibnamefont{{Liebend{\"o}rfer}}},
  \bibinfo{journal}{\aap} \textbf{\bibinfo{volume}{490}}, \bibinfo{pages}{231}
  (\bibinfo{year}{2008}).

\bibitem[{\citenamefont{{Andresen} et~al.}(2016)\citenamefont{{Andresen},
  {Mueller}, {Mueller}, and {Janka}}}]{andresen:16}
\bibinfo{author}{\bibfnamefont{H.}~\bibnamefont{{Andresen}}},
  \bibinfo{author}{\bibfnamefont{B.}~\bibnamefont{{Mueller}}},
  \bibinfo{author}{\bibfnamefont{E.}~\bibnamefont{{Mueller}}},
  \bibnamefont{and} \bibinfo{author}{\bibfnamefont{H.-T.}
  \bibnamefont{{Janka}}}, \bibinfo{journal}{ArXiv e-prints}
  (\bibinfo{year}{2016}).

\bibitem[{\citenamefont{{Kuroda} et~al.}(2014)\citenamefont{{Kuroda},
  {Takiwaki}, and {Kotake}}}]{kuroda:14}
\bibinfo{author}{\bibfnamefont{T.}~\bibnamefont{{Kuroda}}},
  \bibinfo{author}{\bibfnamefont{T.}~\bibnamefont{{Takiwaki}}},
  \bibnamefont{and} \bibinfo{author}{\bibfnamefont{K.}~\bibnamefont{{Kotake}}},
  \bibinfo{journal}{\prd} \textbf{\bibinfo{volume}{89}}, \bibinfo{eid}{044011}
  (\bibinfo{year}{2014}).

\bibitem[{\citenamefont{{Kotake} et~al.}(2011)\citenamefont{{Kotake},
  {Iwakami-Nakano}, and {Ohnishi}}}]{kotake:11}
\bibinfo{author}{\bibfnamefont{K.}~\bibnamefont{{Kotake}}},
  \bibinfo{author}{\bibfnamefont{W.}~\bibnamefont{{Iwakami-Nakano}}},
  \bibnamefont{and}
  \bibinfo{author}{\bibfnamefont{N.}~\bibnamefont{{Ohnishi}}},
  \bibinfo{journal}{\apj} \textbf{\bibinfo{volume}{736}}, \bibinfo{eid}{124}
  (\bibinfo{year}{2011}).

\bibitem[{\citenamefont{{Kotake} et~al.}(2009)\citenamefont{{Kotake},
  {Iwakami}, {Ohnishi}, and {Yamada}}}]{kotake:09}
\bibinfo{author}{\bibfnamefont{K.}~\bibnamefont{{Kotake}}},
  \bibinfo{author}{\bibfnamefont{W.}~\bibnamefont{{Iwakami}}},
  \bibinfo{author}{\bibfnamefont{N.}~\bibnamefont{{Ohnishi}}},
  \bibnamefont{and} \bibinfo{author}{\bibfnamefont{S.}~\bibnamefont{{Yamada}}},
  \bibinfo{journal}{\apjl} \textbf{\bibinfo{volume}{697}},
  \bibinfo{pages}{L133} (\bibinfo{year}{2009}).

\bibitem[{\citenamefont{{Yakunin} et~al.}(2014)\citenamefont{{Yakunin},
  {Marronetti}, {Messer}, {Mezzacappa}, {Lentz}, {Bruenn}, {Hix}, {Harris}, and
  {Blondin}}}]{yakunin:14}
\bibinfo{author}{\bibfnamefont{K.}~\bibnamefont{{Yakunin}}},
  \bibinfo{author}{\bibfnamefont{P.}~\bibnamefont{{Marronetti}}},
  \bibinfo{author}{\bibfnamefont{O.~B.} \bibnamefont{{Messer}}},
  \bibinfo{author}{\bibfnamefont{A.}~\bibnamefont{{Mezzacappa}}},
  \bibinfo{author}{\bibfnamefont{E.~J.} \bibnamefont{{Lentz}}},
  \bibinfo{author}{\bibfnamefont{S.~W.} \bibnamefont{{Bruenn}}},
  \bibinfo{author}{\bibfnamefont{W.~R.} \bibnamefont{{Hix}}},
  \bibinfo{author}{\bibfnamefont{J.~A.} \bibnamefont{{Harris}}},
  \bibnamefont{and} \bibinfo{author}{\bibfnamefont{J.~M.}
  \bibnamefont{{Blondin}}}, in \emph{\bibinfo{booktitle}{American Astronomical
  Society Meeting Abstracts \#223}} (\bibinfo{year}{2014}), vol.
  \bibinfo{volume}{223} of \emph{\bibinfo{series}{American Astronomical Society
  Meeting Abstracts}}, p. \bibinfo{pages}{354.09}.

\bibitem[{\citenamefont{{Weinberg} and {Quataert}}(2008)}]{weinberg:08}
\bibinfo{author}{\bibfnamefont{N.~N.} \bibnamefont{{Weinberg}}}
  \bibnamefont{and}
  \bibinfo{author}{\bibfnamefont{E.}~\bibnamefont{{Quataert}}},
  \bibinfo{journal}{\mnras} \textbf{\bibinfo{volume}{387}},
  \bibinfo{pages}{L64} (\bibinfo{year}{2008}).

\bibitem[{\citenamefont{{Murphy} et~al.}(2009)\citenamefont{{Murphy}, {Ott},
  and {Burrows}}}]{murphy:09}
\bibinfo{author}{\bibfnamefont{J.~W.} \bibnamefont{{Murphy}}},
  \bibinfo{author}{\bibfnamefont{C.~D.} \bibnamefont{{Ott}}}, \bibnamefont{and}
  \bibinfo{author}{\bibfnamefont{A.}~\bibnamefont{{Burrows}}},
  \bibinfo{journal}{\apj} \textbf{\bibinfo{volume}{707}}, \bibinfo{pages}{1173}
  (\bibinfo{year}{2009}).

\bibitem[{\citenamefont{{Dimmelmeier} et~al.}(2008)\citenamefont{{Dimmelmeier},
  {Ott}, {Marek}, and {Janka}}}]{dimmelmeier:08}
\bibinfo{author}{\bibfnamefont{H.}~\bibnamefont{{Dimmelmeier}}},
  \bibinfo{author}{\bibfnamefont{C.~D.} \bibnamefont{{Ott}}},
  \bibinfo{author}{\bibfnamefont{A.}~\bibnamefont{{Marek}}}, \bibnamefont{and}
  \bibinfo{author}{\bibfnamefont{H.-T.} \bibnamefont{{Janka}}},
  \bibinfo{journal}{\prd} \textbf{\bibinfo{volume}{78}},
  \bibinfo{pages}{064056} (\bibinfo{year}{2008}).

\bibitem[{\citenamefont{{Ott} et~al.}(2006)\citenamefont{{Ott}, {Burrows},
  {Thompson}, {Livne}, and {Walder}}}]{ott:06spin}
\bibinfo{author}{\bibfnamefont{C.~D.} \bibnamefont{{Ott}}},
  \bibinfo{author}{\bibfnamefont{A.}~\bibnamefont{{Burrows}}},
  \bibinfo{author}{\bibfnamefont{T.~A.} \bibnamefont{{Thompson}}},
  \bibinfo{author}{\bibfnamefont{E.}~\bibnamefont{{Livne}}}, \bibnamefont{and}
  \bibinfo{author}{\bibfnamefont{R.}~\bibnamefont{{Walder}}},
  \bibinfo{journal}{\apjs} \textbf{\bibinfo{volume}{164}}, \bibinfo{pages}{130}
  (\bibinfo{year}{2006}).

\bibitem[{\citenamefont{{Takiwaki} and {Kotake}}(2011)}]{takiwaki:11}
\bibinfo{author}{\bibfnamefont{T.}~\bibnamefont{{Takiwaki}}} \bibnamefont{and}
  \bibinfo{author}{\bibfnamefont{K.}~\bibnamefont{{Kotake}}},
  \bibinfo{journal}{\apj} \textbf{\bibinfo{volume}{743}}, \bibinfo{eid}{30}
  (\bibinfo{year}{2011}).

\bibitem[{\citenamefont{{Heger} et~al.}(2005)\citenamefont{{Heger}, {Woosley},
  and {Spruit}}}]{heger:05}
\bibinfo{author}{\bibfnamefont{A.}~\bibnamefont{{Heger}}},
  \bibinfo{author}{\bibfnamefont{S.~E.} \bibnamefont{{Woosley}}},
  \bibnamefont{and} \bibinfo{author}{\bibfnamefont{H.~C.}
  \bibnamefont{{Spruit}}}, \bibinfo{journal}{\apj}
  \textbf{\bibinfo{volume}{626}}, \bibinfo{pages}{350} (\bibinfo{year}{2005}).

\bibitem[{\citenamefont{{Rembiasz} et~al.}(2016)\citenamefont{{Rembiasz},
  {Guilet}, {Obergaulinger}, {Cerd{\'a}-Dur{\'a}n}, {Aloy}, and
  {M{\"u}ller}}}]{2016MNRAS.460.3316R}
\bibinfo{author}{\bibfnamefont{T.}~\bibnamefont{{Rembiasz}}},
  \bibinfo{author}{\bibfnamefont{J.}~\bibnamefont{{Guilet}}},
  \bibinfo{author}{\bibfnamefont{M.}~\bibnamefont{{Obergaulinger}}},
  \bibinfo{author}{\bibfnamefont{P.}~\bibnamefont{{Cerd{\'a}-Dur{\'a}n}}},
  \bibinfo{author}{\bibfnamefont{M.~A.} \bibnamefont{{Aloy}}},
  \bibnamefont{and}
  \bibinfo{author}{\bibfnamefont{E.}~\bibnamefont{{M{\"u}ller}}},
  \bibinfo{journal}{\mnras} \textbf{\bibinfo{volume}{460}},
  \bibinfo{pages}{3316} (\bibinfo{year}{2016}).

\bibitem[{\citenamefont{{M{\"o}sta} et~al.}(2015)\citenamefont{{M{\"o}sta},
  {Ott}, {Radice}, {Roberts}, {Schnetter}, and {Haas}}}]{2015Natur.528..376M}
\bibinfo{author}{\bibfnamefont{P.}~\bibnamefont{{M{\"o}sta}}},
  \bibinfo{author}{\bibfnamefont{C.~D.} \bibnamefont{{Ott}}},
  \bibinfo{author}{\bibfnamefont{D.}~\bibnamefont{{Radice}}},
  \bibinfo{author}{\bibfnamefont{L.~F.} \bibnamefont{{Roberts}}},
  \bibinfo{author}{\bibfnamefont{E.}~\bibnamefont{{Schnetter}}},
  \bibnamefont{and} \bibinfo{author}{\bibfnamefont{R.}~\bibnamefont{{Haas}}},
  \bibinfo{journal}{\nat} \textbf{\bibinfo{volume}{528}}, \bibinfo{pages}{376}
  (\bibinfo{year}{2015}).

\bibitem[{\citenamefont{{Abdikamalov} et~al.}(2014)\citenamefont{{Abdikamalov},
  {Gossan}, {DeMaio}, and {Ott}}}]{abdikamalov:14}
\bibinfo{author}{\bibfnamefont{E.}~\bibnamefont{{Abdikamalov}}},
  \bibinfo{author}{\bibfnamefont{S.}~\bibnamefont{{Gossan}}},
  \bibinfo{author}{\bibfnamefont{A.~M.} \bibnamefont{{DeMaio}}},
  \bibnamefont{and} \bibinfo{author}{\bibfnamefont{C.~D.} \bibnamefont{{Ott}}},
  \bibinfo{journal}{\prd} \textbf{\bibinfo{volume}{90}}, \bibinfo{eid}{044001}
  (\bibinfo{year}{2014}).

\bibitem[{\citenamefont{{Scheidegger}
  et~al.}(2010{\natexlab{a}})\citenamefont{{Scheidegger}, {Whitehouse},
  {K{\"a}ppeli}, and {Liebend{\"o}rfer}}}]{scheidegger:10}
\bibinfo{author}{\bibfnamefont{S.}~\bibnamefont{{Scheidegger}}},
  \bibinfo{author}{\bibfnamefont{S.~C.} \bibnamefont{{Whitehouse}}},
  \bibinfo{author}{\bibfnamefont{R.}~\bibnamefont{{K{\"a}ppeli}}},
  \bibnamefont{and}
  \bibinfo{author}{\bibfnamefont{M.}~\bibnamefont{{Liebend{\"o}rfer}}},
  \bibinfo{journal}{\cqg} \textbf{\bibinfo{volume}{27}},
  \bibinfo{pages}{114101} (\bibinfo{year}{2010}{\natexlab{a}}).

\bibitem[{\citenamefont{{Rampp} et~al.}(1998)\citenamefont{{Rampp}, {Mueller},
  and {Ruffert}}}]{rampp:98}
\bibinfo{author}{\bibfnamefont{M.}~\bibnamefont{{Rampp}}},
  \bibinfo{author}{\bibfnamefont{E.}~\bibnamefont{{Mueller}}},
  \bibnamefont{and}
  \bibinfo{author}{\bibfnamefont{M.}~\bibnamefont{{Ruffert}}},
  \bibinfo{journal}{\aap} \textbf{\bibinfo{volume}{332}}, \bibinfo{pages}{969}
  (\bibinfo{year}{1998}).

\bibitem[{\citenamefont{{Lattimer} and {Douglas Swesty}}(1991)}]{lattimer:91}
\bibinfo{author}{\bibfnamefont{J.~M.} \bibnamefont{{Lattimer}}}
  \bibnamefont{and} \bibinfo{author}{\bibfnamefont{F.}~\bibnamefont{{Douglas
  Swesty}}}, \bibinfo{journal}{Nuclear Physics A}
  \textbf{\bibinfo{volume}{535}}, \bibinfo{pages}{331} (\bibinfo{year}{1991}).

\bibitem[{\citenamefont{{Shen} et~al.}(1998{\natexlab{a}})\citenamefont{{Shen},
  {Toki}, {Oyamatsu}, and {Sumiyoshi}}}]{shen:98a}
\bibinfo{author}{\bibfnamefont{H.}~\bibnamefont{{Shen}}},
  \bibinfo{author}{\bibfnamefont{H.}~\bibnamefont{{Toki}}},
  \bibinfo{author}{\bibfnamefont{K.}~\bibnamefont{{Oyamatsu}}},
  \bibnamefont{and}
  \bibinfo{author}{\bibfnamefont{K.}~\bibnamefont{{Sumiyoshi}}},
  \bibinfo{journal}{Nuclear Physics A} \textbf{\bibinfo{volume}{637}},
  \bibinfo{pages}{435} (\bibinfo{year}{1998}{\natexlab{a}}).

\bibitem[{\citenamefont{{Shen} et~al.}(1998{\natexlab{b}})\citenamefont{{Shen},
  {Toki}, {Oyamatsu}, and {Sumiyoshi}}}]{shen:98b}
\bibinfo{author}{\bibfnamefont{H.}~\bibnamefont{{Shen}}},
  \bibinfo{author}{\bibfnamefont{H.}~\bibnamefont{{Toki}}},
  \bibinfo{author}{\bibfnamefont{K.}~\bibnamefont{{Oyamatsu}}},
  \bibnamefont{and}
  \bibinfo{author}{\bibfnamefont{K.}~\bibnamefont{{Sumiyoshi}}},
  \bibinfo{journal}{Progress of Theoretical Physics}
  \textbf{\bibinfo{volume}{100}}, \bibinfo{pages}{1013}
  (\bibinfo{year}{1998}{\natexlab{b}}).

\bibitem[{\citenamefont{{Scheidegger}
  et~al.}(2010{\natexlab{b}})\citenamefont{{Scheidegger}, {K{\"a}ppeli},
  {Whitehouse}, {Fischer}, and {Liebend{\"o}rfer}}}]{scheidegger:10b}
\bibinfo{author}{\bibfnamefont{S.}~\bibnamefont{{Scheidegger}}},
  \bibinfo{author}{\bibfnamefont{R.}~\bibnamefont{{K{\"a}ppeli}}},
  \bibinfo{author}{\bibfnamefont{S.~C.} \bibnamefont{{Whitehouse}}},
  \bibinfo{author}{\bibfnamefont{T.}~\bibnamefont{{Fischer}}},
  \bibnamefont{and}
  \bibinfo{author}{\bibfnamefont{M.}~\bibnamefont{{Liebend{\"o}rfer}}},
  \bibinfo{journal}{\aap} \textbf{\bibinfo{volume}{514}}, \bibinfo{pages}{A51}
  (\bibinfo{year}{2010}{\natexlab{b}}).

\bibitem[{\citenamefont{Arnett}(1966)}]{arnett:66}
\bibinfo{author}{\bibfnamefont{W.~D.} \bibnamefont{Arnett}},
  \bibinfo{journal}{Canadian Journal of Physics} \textbf{\bibinfo{volume}{44}},
  \bibinfo{pages}{2553} (\bibinfo{year}{1966}).

\bibitem[{\citenamefont{{Colgate} and {White}}(1966)}]{colgate:66}
\bibinfo{author}{\bibfnamefont{S.~A.} \bibnamefont{{Colgate}}}
  \bibnamefont{and} \bibinfo{author}{\bibfnamefont{R.~H.}
  \bibnamefont{{White}}}, \bibinfo{journal}{\apj}
  \textbf{\bibinfo{volume}{143}}, \bibinfo{pages}{626} (\bibinfo{year}{1966}).

\bibitem[{\citenamefont{{Bethe} and {Wilson}}(1985)}]{bethe:85}
\bibinfo{author}{\bibfnamefont{H.~A.} \bibnamefont{{Bethe}}} \bibnamefont{and}
  \bibinfo{author}{\bibfnamefont{J.~R.} \bibnamefont{{Wilson}}},
  \bibinfo{journal}{\apj} \textbf{\bibinfo{volume}{295}}, \bibinfo{pages}{14}
  (\bibinfo{year}{1985}).

\bibitem[{\citenamefont{{Blondin} et~al.}(2003)\citenamefont{{Blondin},
  {Mezzacappa}, and {DeMarino}}}]{blondin:03}
\bibinfo{author}{\bibfnamefont{J.~M.} \bibnamefont{{Blondin}}},
  \bibinfo{author}{\bibfnamefont{A.}~\bibnamefont{{Mezzacappa}}},
  \bibnamefont{and}
  \bibinfo{author}{\bibfnamefont{C.}~\bibnamefont{{DeMarino}}},
  \bibinfo{journal}{\apj} \textbf{\bibinfo{volume}{584}}, \bibinfo{pages}{971}
  (\bibinfo{year}{2003}).

\bibitem[{\citenamefont{{Foglizzo} et~al.}(2007)\citenamefont{{Foglizzo},
  {Galletti}, {Scheck}, and {Janka}}}]{foglizzo:07}
\bibinfo{author}{\bibfnamefont{T.}~\bibnamefont{{Foglizzo}}},
  \bibinfo{author}{\bibfnamefont{P.}~\bibnamefont{{Galletti}}},
  \bibinfo{author}{\bibfnamefont{L.}~\bibnamefont{{Scheck}}}, \bibnamefont{and}
  \bibinfo{author}{\bibfnamefont{H.-T.} \bibnamefont{{Janka}}},
  \bibinfo{journal}{\apj} \textbf{\bibinfo{volume}{654}}, \bibinfo{pages}{1006}
  (\bibinfo{year}{2007}).

\bibitem[{\citenamefont{{Foglizzo} et~al.}(2006)\citenamefont{{Foglizzo},
  {Scheck}, and {Janka}}}]{foglizzo:06}
\bibinfo{author}{\bibfnamefont{T.}~\bibnamefont{{Foglizzo}}},
  \bibinfo{author}{\bibfnamefont{L.}~\bibnamefont{{Scheck}}}, \bibnamefont{and}
  \bibinfo{author}{\bibfnamefont{H.-T.} \bibnamefont{{Janka}}},
  \bibinfo{journal}{\apj} \textbf{\bibinfo{volume}{652}}, \bibinfo{pages}{1436}
  (\bibinfo{year}{2006}).

\bibitem[{\citenamefont{{Scheck} et~al.}(2008)\citenamefont{{Scheck}, {Janka},
  {Foglizzo}, and {Kifonidis}}}]{scheck:08}
\bibinfo{author}{\bibfnamefont{L.}~\bibnamefont{{Scheck}}},
  \bibinfo{author}{\bibfnamefont{H.-T.} \bibnamefont{{Janka}}},
  \bibinfo{author}{\bibfnamefont{T.}~\bibnamefont{{Foglizzo}}},
  \bibnamefont{and}
  \bibinfo{author}{\bibfnamefont{K.}~\bibnamefont{{Kifonidis}}},
  \bibinfo{journal}{\aap} \textbf{\bibinfo{volume}{477}}, \bibinfo{pages}{931}
  (\bibinfo{year}{2008}).

\bibitem[{\citenamefont{{Misner} et~al.}(1973)\citenamefont{{Misner}, {Thorne},
  and {Wheeler}}}]{mtw}
\bibinfo{author}{\bibfnamefont{C.~W.} \bibnamefont{{Misner}}},
  \bibinfo{author}{\bibfnamefont{K.~S.} \bibnamefont{{Thorne}}},
  \bibnamefont{and} \bibinfo{author}{\bibfnamefont{J.~A.}
  \bibnamefont{{Wheeler}}}, \emph{\bibinfo{title}{{Gravitation}}}
  (\bibinfo{publisher}{San Francisco: W.H.~Freeman and Co.},
  \bibinfo{year}{1973}).

\bibitem[{\citenamefont{{Yakunin} et~al.}(2010)\citenamefont{{Yakunin},
  {Marronetti}, {Mezzacappa}, {Bruenn}, {Lee}, {Chertkow}, {Hix}, {Blondin},
  {Lentz}, {Bronson Messer} et~al.}}]{yakunin:10}
\bibinfo{author}{\bibfnamefont{K.~N.} \bibnamefont{{Yakunin}}},
  \bibinfo{author}{\bibfnamefont{P.}~\bibnamefont{{Marronetti}}},
  \bibinfo{author}{\bibfnamefont{A.}~\bibnamefont{{Mezzacappa}}},
  \bibinfo{author}{\bibfnamefont{S.~W.} \bibnamefont{{Bruenn}}},
  \bibinfo{author}{\bibfnamefont{C.-T.} \bibnamefont{{Lee}}},
  \bibinfo{author}{\bibfnamefont{M.~A.} \bibnamefont{{Chertkow}}},
  \bibinfo{author}{\bibfnamefont{W.~R.} \bibnamefont{{Hix}}},
  \bibinfo{author}{\bibfnamefont{J.~M.} \bibnamefont{{Blondin}}},
  \bibinfo{author}{\bibfnamefont{E.~J.} \bibnamefont{{Lentz}}},
  \bibinfo{author}{\bibfnamefont{O.~E.} \bibnamefont{{Bronson Messer}}},
  \bibnamefont{et~al.}, \bibinfo{journal}{Class. Quantum Grav.}
  \textbf{\bibinfo{volume}{27}}, \bibinfo{pages}{194005}
  (\bibinfo{year}{2010}).

\bibitem[{\citenamefont{{M{\"u}ller} et~al.}(2012)\citenamefont{{M{\"u}ller},
  {Janka}, and {Wongwathanarat}}}]{mueller:e12}
\bibinfo{author}{\bibfnamefont{E.}~\bibnamefont{{M{\"u}ller}}},
  \bibinfo{author}{\bibfnamefont{H.-T.} \bibnamefont{{Janka}}},
  \bibnamefont{and}
  \bibinfo{author}{\bibfnamefont{A.}~\bibnamefont{{Wongwathanarat}}},
  \bibinfo{journal}{\aap} \textbf{\bibinfo{volume}{537}}, \bibinfo{eid}{A63}
  (\bibinfo{year}{2012}).

\bibitem[{\citenamefont{Pürrer}(2014)}]{0264-9381-31-19-195010}
\bibinfo{author}{\bibfnamefont{M.}~\bibnamefont{Pürrer}},
  \bibinfo{journal}{Classical and Quantum Gravity}
  \textbf{\bibinfo{volume}{31}}, \bibinfo{pages}{195010}
  (\bibinfo{year}{2014}),
  \urlprefix\url{http://stacks.iop.org/0264-9381/31/i=19/a=195010}.

\bibitem[{\citenamefont{Blackman et~al.}(2015)\citenamefont{Blackman, Field,
  Galley, Szil\'agyi, Scheel, Tiglio, and Hemberger}}]{PhysRevLett.115.121102}
\bibinfo{author}{\bibfnamefont{J.}~\bibnamefont{Blackman}},
  \bibinfo{author}{\bibfnamefont{S.~E.} \bibnamefont{Field}},
  \bibinfo{author}{\bibfnamefont{C.~R.} \bibnamefont{Galley}},
  \bibinfo{author}{\bibfnamefont{B.}~\bibnamefont{Szil\'agyi}},
  \bibinfo{author}{\bibfnamefont{M.~A.} \bibnamefont{Scheel}},
  \bibinfo{author}{\bibfnamefont{M.}~\bibnamefont{Tiglio}}, \bibnamefont{and}
  \bibinfo{author}{\bibfnamefont{D.~A.} \bibnamefont{Hemberger}},
  \bibinfo{journal}{Phys. Rev. Lett.} \textbf{\bibinfo{volume}{115}},
  \bibinfo{pages}{121102} (\bibinfo{year}{2015}),
  \urlprefix\url{http://link.aps.org/doi/10.1103/PhysRevLett.115.121102}.

\bibitem[{\citenamefont{Clark et~al.}(2016)\citenamefont{Clark, Bauswein,
  Stergioulas, and Shoemaker}}]{0264-9381-33-8-085003}
\bibinfo{author}{\bibfnamefont{J.~A.} \bibnamefont{Clark}},
  \bibinfo{author}{\bibfnamefont{A.}~\bibnamefont{Bauswein}},
  \bibinfo{author}{\bibfnamefont{N.}~\bibnamefont{Stergioulas}},
  \bibnamefont{and}
  \bibinfo{author}{\bibfnamefont{D.}~\bibnamefont{Shoemaker}},
  \bibinfo{journal}{Classical and Quantum Gravity}
  \textbf{\bibinfo{volume}{33}}, \bibinfo{pages}{085003}
  (\bibinfo{year}{2016}),
  \urlprefix\url{http://stacks.iop.org/0264-9381/33/i=8/a=085003}.

\bibitem[{\citenamefont{{Powell} et~al.}(2015)\citenamefont{{Powell},
  {Trifir{\`o}}, {Cuoco}, {Heng}, and {Cavagli{\`a}}}}]{powell:15}
\bibinfo{author}{\bibfnamefont{J.}~\bibnamefont{{Powell}}},
  \bibinfo{author}{\bibfnamefont{D.}~\bibnamefont{{Trifir{\`o}}}},
  \bibinfo{author}{\bibfnamefont{E.}~\bibnamefont{{Cuoco}}},
  \bibinfo{author}{\bibfnamefont{I.~S.} \bibnamefont{{Heng}}},
  \bibnamefont{and}
  \bibinfo{author}{\bibfnamefont{M.}~\bibnamefont{{Cavagli{\`a}}}},
  \bibinfo{journal}{\cqg} \textbf{\bibinfo{volume}{32}}, \bibinfo{eid}{215012}
  (\bibinfo{year}{2015}).

\bibitem[{\citenamefont{{Powell} et~al.}(2016)\citenamefont{{Powell},
  {Torres-Forn{\'e}}, {Lynch}, {Trifir{\`o}}, {Cuoco}, {Cavagli{\`a}}, {Heng},
  and {Font}}}]{powell:16}
\bibinfo{author}{\bibfnamefont{J.}~\bibnamefont{{Powell}}},
  \bibinfo{author}{\bibfnamefont{A.}~\bibnamefont{{Torres-Forn{\'e}}}},
  \bibinfo{author}{\bibfnamefont{R.}~\bibnamefont{{Lynch}}},
  \bibinfo{author}{\bibfnamefont{D.}~\bibnamefont{{Trifir{\`o}}}},
  \bibinfo{author}{\bibfnamefont{E.}~\bibnamefont{{Cuoco}}},
  \bibinfo{author}{\bibfnamefont{M.}~\bibnamefont{{Cavagli{\`a}}}},
  \bibinfo{author}{\bibfnamefont{I.~S.} \bibnamefont{{Heng}}},
  \bibnamefont{and} \bibinfo{author}{\bibfnamefont{J.~A.}
  \bibnamefont{{Font}}}, \bibinfo{journal}{ArXiv e-prints}
  (\bibinfo{year}{2016}).

\bibitem[{LAL()}]{LAL}
\emph{\bibinfo{title}{Ligo algorithm library}},
  \urlprefix\url{www.lsc-group.phys.uwm.edu/daswg/projects/lalsuite.html}.

\bibitem[{\citenamefont{{Veitch} et~al.}(2015)\citenamefont{{Veitch},
  {Raymond}, {Farr}, {Farr}, {Graff}, {Vitale}, {Aylott}, {Blackburn},
  {Christensen}, {Coughlin} et~al.}}]{veitch:15}
\bibinfo{author}{\bibfnamefont{J.}~\bibnamefont{{Veitch}}},
  \bibinfo{author}{\bibfnamefont{V.}~\bibnamefont{{Raymond}}},
  \bibinfo{author}{\bibfnamefont{B.}~\bibnamefont{{Farr}}},
  \bibinfo{author}{\bibfnamefont{W.}~\bibnamefont{{Farr}}},
  \bibinfo{author}{\bibfnamefont{P.}~\bibnamefont{{Graff}}},
  \bibinfo{author}{\bibfnamefont{S.}~\bibnamefont{{Vitale}}},
  \bibinfo{author}{\bibfnamefont{B.}~\bibnamefont{{Aylott}}},
  \bibinfo{author}{\bibfnamefont{K.}~\bibnamefont{{Blackburn}}},
  \bibinfo{author}{\bibfnamefont{N.}~\bibnamefont{{Christensen}}},
  \bibinfo{author}{\bibfnamefont{M.}~\bibnamefont{{Coughlin}}},
  \bibnamefont{et~al.}, \bibinfo{journal}{\prd} \textbf{\bibinfo{volume}{91}},
  \bibinfo{eid}{042003} (\bibinfo{year}{2015}).

\bibitem[{\citenamefont{{Essick} et~al.}(2015)\citenamefont{{Essick}, {Vitale},
  {Katsavounidis}, {Vedovato}, and {Klimenko}}}]{essick:15}
\bibinfo{author}{\bibfnamefont{R.}~\bibnamefont{{Essick}}},
  \bibinfo{author}{\bibfnamefont{S.}~\bibnamefont{{Vitale}}},
  \bibinfo{author}{\bibfnamefont{E.}~\bibnamefont{{Katsavounidis}}},
  \bibinfo{author}{\bibfnamefont{G.}~\bibnamefont{{Vedovato}}},
  \bibnamefont{and}
  \bibinfo{author}{\bibfnamefont{S.}~\bibnamefont{{Klimenko}}},
  \bibinfo{journal}{\apj} \textbf{\bibinfo{volume}{800}}, \bibinfo{eid}{81}
  (\bibinfo{year}{2015}).

\bibitem[{\citenamefont{{Lynch} et~al.}(2015)\citenamefont{{Lynch}, {Vitale},
  {Essick}, {Katsavounidis}, and {Robinet}}}]{lynch:15}
\bibinfo{author}{\bibfnamefont{R.}~\bibnamefont{{Lynch}}},
  \bibinfo{author}{\bibfnamefont{S.}~\bibnamefont{{Vitale}}},
  \bibinfo{author}{\bibfnamefont{R.}~\bibnamefont{{Essick}}},
  \bibinfo{author}{\bibfnamefont{E.}~\bibnamefont{{Katsavounidis}}},
  \bibnamefont{and}
  \bibinfo{author}{\bibfnamefont{F.}~\bibnamefont{{Robinet}}},
  \bibinfo{journal}{arXiv:1511.05955}  (\bibinfo{year}{2015}).

\bibitem[{\citenamefont{Sivia}(1996)}]{sivia:96}
\bibinfo{author}{\bibfnamefont{D.~S.} \bibnamefont{Sivia}},
  \emph{\bibinfo{title}{Data Analysis, A Bayesian Tutorial}}
  (\bibinfo{publisher}{Oxford}, \bibinfo{year}{1996}).

\bibitem[{\citenamefont{T.~Hastie}(2001)}]{PCA-clustering}
\bibinfo{author}{\bibfnamefont{J.~F.} \bibnamefont{T.~Hastie},
  \bibfnamefont{R.~Tibshirani}}, \emph{\bibinfo{title}{The Elements of
  Statistical Learning}} (\bibinfo{publisher}{Springer}, \bibinfo{year}{2001}).

\bibitem[{\citenamefont{{Vallisneri} et~al.}(2015)}]{vallisneri:15}
\bibinfo{author}{\bibfnamefont{M.}~\bibnamefont{{Vallisneri}}}
  \bibnamefont{et~al.}, \bibinfo{journal}{Journal of Physics Conference Series}
  \textbf{\bibinfo{volume}{610}}, \bibinfo{eid}{012021} (\bibinfo{year}{2015}).

\bibitem[{\citenamefont{{Yakunin} et~al.}(2015)}]{yakunin:15}
\bibinfo{author}{\bibfnamefont{K.~N.} \bibnamefont{{Yakunin}}}
  \bibnamefont{et~al.}, \bibinfo{journal}{\prd} \textbf{\bibinfo{volume}{92}},
  \bibinfo{eid}{084040} (\bibinfo{year}{2015}).

\bibitem[{\citenamefont{{M{\"u}ller} et~al.}(2013)\citenamefont{{M{\"u}ller},
  {Janka}, and {Marek}}}]{muller:13}
\bibinfo{author}{\bibfnamefont{B.}~\bibnamefont{{M{\"u}ller}}},
  \bibinfo{author}{\bibfnamefont{H.-T.} \bibnamefont{{Janka}}},
  \bibnamefont{and} \bibinfo{author}{\bibfnamefont{A.}~\bibnamefont{{Marek}}},
  \bibinfo{journal}{\apj} \textbf{\bibinfo{volume}{766}}, \bibinfo{eid}{43}
  (\bibinfo{year}{2013}).

\bibitem[{\citenamefont{Cornish and Littenberg}(2015)}]{bayeswave}
\bibinfo{author}{\bibfnamefont{N.~J.} \bibnamefont{Cornish}} \bibnamefont{and}
  \bibinfo{author}{\bibfnamefont{T.~B.} \bibnamefont{Littenberg}},
  \bibinfo{journal}{Classical and Quantum Gravity}
  \textbf{\bibinfo{volume}{32}}, \bibinfo{pages}{135012}
  (\bibinfo{year}{2015}),
  \urlprefix\url{http://stacks.iop.org/0264-9381/32/i=13/a=135012}.

\bibitem[{\citenamefont{{Klimenko} et~al.}(2016)}]{2016PhRvD..93d2004K}
\bibinfo{author}{\bibfnamefont{S.}~\bibnamefont{{Klimenko}}}
  \bibnamefont{et~al.}, \bibinfo{journal}{\prd} \textbf{\bibinfo{volume}{93}},
  \bibinfo{eid}{042004} (\bibinfo{year}{2016}).

\bibitem[{\citenamefont{Klimenko et~al.}(2008)}]{0264-9381-25-11-114029}
\bibinfo{author}{\bibfnamefont{S.}~\bibnamefont{Klimenko}}
  \bibnamefont{et~al.}, \bibinfo{journal}{Classical and Quantum Gravity}
  \textbf{\bibinfo{volume}{25}}, \bibinfo{pages}{114029}
  (\bibinfo{year}{2008}),
  \urlprefix\url{http://stacks.iop.org/0264-9381/25/i=11/a=114029}.

\end{thebibliography}

\end{document}